\documentclass[aps,prx,amsmath,amssymb,nofootinbib,longbibliography,superscriptaddress,onecolumn,11pt]{revtex4-2}

\usepackage[english]{babel}
\usepackage{graphicx}
\usepackage{bm}
\usepackage{float}
\usepackage[svgnames]{xcolor}
\usepackage{booktabs}
\usepackage{comment}
\usepackage{url}
\usepackage[colorlinks, linkcolor=NavyBlue, citecolor=NavyBlue, urlcolor=NavyBlue]{hyperref}

\newcommand{\Tq}{T_{\mathrm{quant}}}
\newcommand{\Tcoh}{T_{\mathrm{coh}}}
\newcommand{\Tfs}{T_{\mathrm{fs}}}
\newcommand{\chic}{\chi_c}
\newcommand{\chis}{\chi_s}
\newcommand{\Dc}{D_c}
\newcommand{\sdc}{\sigma_{\mathrm{dc}}}

\newcommand{\DFcage}{\Delta F_{\mathrm{cage}}}
\newcommand{\xicage}{\xi_{\mathrm{cage}}}
\newcommand{\Ncage}{N_{\mathrm{cage}}}

\newcommand{\figplaceholder}[2]{%
  \fbox{\parbox[c][#1][c]{0.92\linewidth}{\centering
    \textbf{FIGURE PLACEHOLDER}\\[2pt] \small #2}}}

\begin{document}

\title{Bad metallicity in the semi-quantum regime of the Hubbard model}

\author{Evyatar Tulipman}
\email{tulipman@stanford.edu}
\affiliation{Geballe Laboratory for Advanced Materials, Stanford University, Stanford, CA 94305, USA}
\affiliation{Department of Physics, Stanford University, Stanford, CA 94305, USA}

\author{Vadim Oganesyan}
\affiliation{Physics Program and Initiative for the Theoretical Sciences, The Graduate Center, CUNY, New York, NY 10016, USA}
\affiliation{Department of Physics and Astronomy, College of Staten Island, CUNY, Staten Island, NY 10314, USA}

\author{Thomas P. Devereaux}
\affiliation{Geballe Laboratory for Advanced Materials, Stanford University, Stanford, CA 94305, USA}
\affiliation{Stanford Institute for Materials and Energy Sciences, SLAC National Accelerator Laboratory, Menlo Park, CA 94025, USA}
\affiliation{Department of Materials Science and Engineering, Stanford University, Stanford, CA 94305, USA}

\author{Steven A. Kivelson}
\affiliation{Department of Physics, Stanford University, Stanford, CA 94305, USA}

\date{\today}

\begin{abstract}
Bad metals exhibit approximately $T$-linear dc resistivity beyond the
Ioffe--Regel limit. That this behavior occurs in systems with radically
different ground states suggests it is a generic manifestation of strong
local correlations. We test this hypothesis in the infinite-$U$ Hubbard
model with small hole densities using exact diagonalization to compute
thermodynamic and transport properties of finite clusters. Upon cooling,
we find an intermediate temperature range, an electronic analogue of the
``semi-quantum regime'' of liquid helium, in which quantum effects produce
a roughly $T$-independent compressibility, yet the resistivity is $T$-linear
and exceeds the Ioffe--Regel limit.
Entry into this regime is accompanied by the formation of quasi-local
ferromagnetic ``spin cages'' around doped holes, regions that facilitate
local quantum motion embedded in a fluctuating spin background, analogous
to the transient crystalline cages thought to control incoherent transport
in semi-quantum liquid helium.
Remarkably, despite the simplicity of the model, the bad metal behavior found
here resembles that seen in various material platforms.
\end{abstract}

\maketitle


\section{Introduction}
\label{sec:intro}

In an ordinary metal, charge transport is dominated by electronic quasiparticles with a mean free path $\ell$ much longer than the Fermi wavelength $\lambda_F$. This quasiparticle picture breaks down near the Ioffe--Regel limit, $\ell\sim\lambda_F$~\cite{IoffeRegel1960}, where conventional metals often show resistivity saturation~\cite{Gunnarsson2003}. Bad metals violate this expectation:
their dc resistivity, $\rho$, stays approximately linear in temperature and grows
past the Ioffe--Regel scale without saturating
\cite{EmeryKivelson1995,Hussey2004,Phillips2022}. Such behavior appears across
strongly correlated systems from the cuprates \cite{Keimer2015} to
infinite-layer nickelates \cite{Lee2023,Hsu2024}, iron-based
\cite{Kasahara2010} and organic
conductors \cite{BlochCowanPoehler1974}, alkali-doped fullerides \cite{Gunnarsson2003},
ruthenates \cite{Tyler1998}, cold-atom Fermi--Hubbard systems
\cite{Brown2019} and possibly also magic-angle graphene~\cite{Cao2020TBGStrangeMetal}.
While these materials exhibit extremely diverse low-temperature ($T$) ordering tendencies,
at more elevated $T$ they display similar bad-metallic transport,
suggesting that bad metallicity is a generic finite-temperature regime of
strongly correlated electrons rather than a property of any particular
ordered state, quantum critical regime, or any other material-specific mechanism.

This motivates the central question: can strong local correlations alone produce bad metallicity, without invoking disorder or phonons,
and if so, what is the nature of the resulting bad-metal state?
Specifically, we will focus on temperatures above any ordering transition but below the
electronic bandwidth.

Hubbard and Hubbard-like models have long provided the standard microscopic language for strongly correlated electronic systems \cite{ArovasBergKivelsonRaghu2022,Dagotto1994,LeeNagaosaWen2006,QinSchaferAndergassenCorbozGull2022}. Prior studies have shown that they can capture aspects of bad-metal phenomenology~\cite{JaklicPrelovsek1994Conductivity,JaklicPrelovsek1995ChargeDynamics,ZemljicPrelovsek2005Transport,Huang2019,Zhao2025Emery,Deng2013,MerinoMcKenzie2000,Vucicevic2015,Vranic2020,JaklicPrelovsek2000, Kokalj2017, LindnerAuerbach2010,Liu2026Transport}, but these studies typically focus on specific
material-motivated parameters, where the existence of
mesoscale ordering tendencies can obscure the origin of bad metallicity.
Note that in models with a bounded single-particle spectrum, there is a well characterized ``high temperature regime,''
$T \gg T_{ \rm quant}$, where
$T_{ \rm  quant}$ is a
crossover scale
proportional to the bare bandwidth,
above which there is also
an approximately
$T$-linear resistivity,
but one that arises from a $T$-independent diffusion constant, $D_c$, and a
compressibility,
$\chi_c \sim1/T$ \cite{MukerjeeOganesyanHuse2006,Perepelitsky2016, Kokalj2017}.
However, this high $T$ analysis  breaks down at the
temperatures
below
the bandwidth that are relevant to the problem of bad metallicity in most realistic scenarios.

\section{The semi-quantum regime}
\label{sec:semiquantum}

\begin{figure}[t]
\centering
\includegraphics[width=0.5\linewidth]{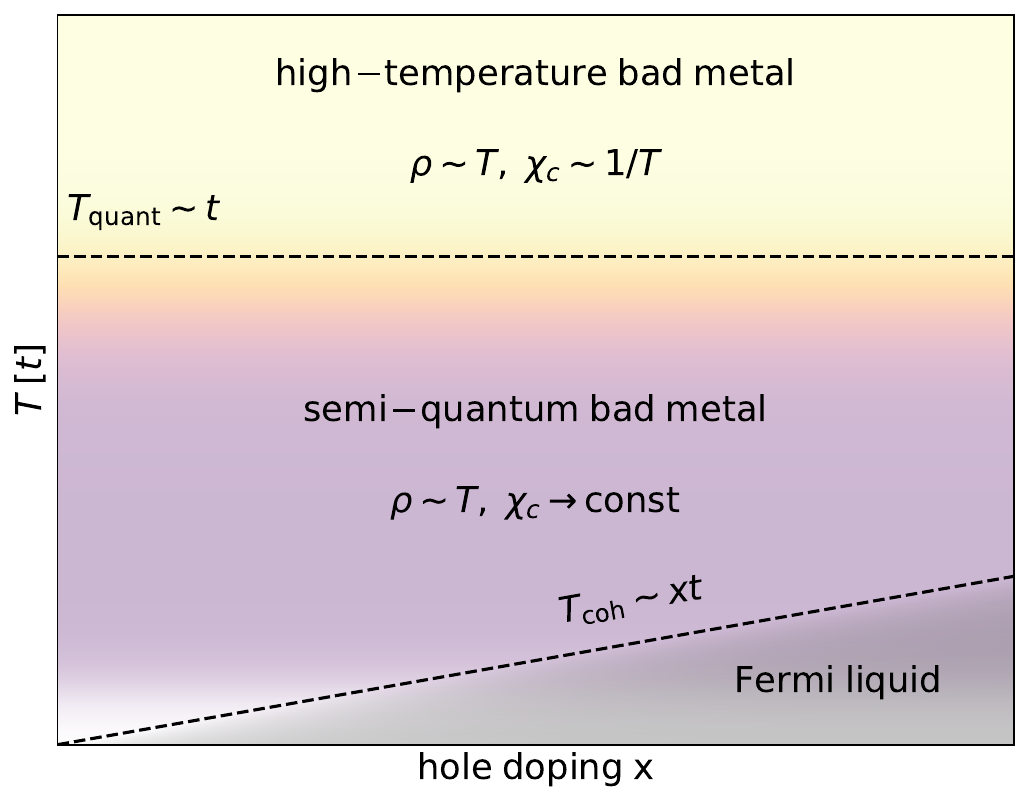}
\caption{\textbf{Schematic transport regimes of the doped infinite-$U$ Hubbard model.}
The semi-quantum bad metal is the
intermediate regime between the high-temperature bad metal, $T\gtrsim T_{\rm quant}\sim t$, and the low-temperature coherent regime, $T\lesssim T_{\rm coh}\sim xt$, where Fermi-liquid behavior is expected to emerge.
Both bad-metal regimes exhibit approximately $T$-linear resistivity above the Ioffe--Regel scale.
In the high-$T$ regime this behavior is tied to
a $T$-dependent
compressibility, $\chi_c\sim 1/T$, whereas in the semi-quantum regime $\chi_c$ departs from this form and approaches a constant, signaling the onset of quantum dynamics, even as charge transport remains incoherent.
Dashed lines indicate crossovers, not sharp phase boundaries.
}
\label{fig:phase_diagram_schematic}
\end{figure}

The infinite-$U$ Hubbard model
is an elegant  setting
in which to investigate the effects of strong local correlations:
It has a single energy scale, the
hopping matrix element, $t$;
instead of an interaction there is simply  the local Hilbert space constraint of no double occupancy.
We study it on the two-dimensional square lattice,
\begin{align}
 H=-t\sum_{\langle ij\rangle,\sigma}(\tilde c^\dagger_{i\sigma}\tilde
c_{j\sigma}+\mathrm{h.c.})
\end{align}
where
$\tilde c_{i\sigma}=c_{i\sigma}(1-n_{i\bar\sigma})$
and $n_{i\bar\sigma}$ is the electron density with spin polarization $\bar{\sigma}$ on site $i$.
At half filling, $\langle n_{i\sigma}\rangle = 1/2$, it is
a Mott insulator by construction; away from half filling it has no
finite-temperature ordered phase.
The model  exhibits an emergent low-energy coherence scale, $T_{\rm coh} $,  which in the limit of small hole concentration, $x\ll 1$, is parametrically small, $T_{\rm coh} \sim xt \ll T_{\rm quant} $. More concretely,
the ground state is ferromagnetic for a single hole
\cite{Nagaoka1966} and was demonstrated numerically to remain so up to hole-doping $x\lesssim0.2$ in two dimensions
\cite{LiuYaoBergWhiteKivelson2012}, with
a spin-stiffness scale
$\rho_s\sim xt$
at $T=0$.  Thus,
$T_{\rm coh}\sim xt$ can be taken to be the
temperature below which the ferromagnetic correlation length begins to grow exponentially (i.e.  $\xi\sim e^{2\pi\rho_s/T}$) with decreasing $T$
(App.~\ref{sec:stiffness}).

Here, we characterize the intermediate regime
$T_{\rm quant}\gtrsim T \gtrsim T_{\rm coh}$, schematically shown in
Fig.~\ref{fig:phase_diagram_schematic}, where $T_{\rm quant}\sim t$ marks the onset of quantum dynamics and
$T_{\rm coh}\sim xt$ the
scale
at which coherent charge motion onsets. Using exact
diagonalization, we obtain the full many-body spectrum and all current matrix
elements, enabling numerically exact calculations of thermodynamics and
transport on $4\times4$, $\sqrt{18}\times\sqrt{18}$, and $4\times5$
clusters. We find that, upon cooling
from above $T_{\rm quant}$,
where the resistivity is already $T$-linear but the compressibility $\sim 1/T$, the system enters a
regime in which thermodynamic quantities such as the compressibility
reflect the quantum dynamics, yet the resistivity remains
above the Ioffe--Regel limit and approximately $T$-linear,
albeit with a
different slope than at high-T. The essential
hierarchy,
$T_{\rm quant}\gg T_{\rm coh}$,
has no analogue in a weakly correlated
metal.

Such a separation between the scales
is the defining feature of the ``semi-quantum liquid'' introduced by Andreev and Kosevich~\cite{Andreev1978,AndreevKosevich1979} in the context of liquid He.  Their key physical insight
was that in a range of $T$, quantum effects are manifest when the level spacings associated with the motion of a single particle  in a local ``cage'' are larger than $T$, while coherent
exchange processes that are consequences of quantum statistics are only significant at much lower $T$.  They further argued that this regime exhibits a viscosity (and hence a diffusion constant) proportional to $1/T$.
In the present case, the local cage is not produced by positional correlations, as in liquid helium, but by the fluctuating spin background through which a doped hole moves. As we show below, the hole develops a finite ferromagnetic environment in which its motion is less kinetically frustrated.
This polarized region acts as a self-generated ``spin cage'' that permits
quantum motion locally while inhibiting coherent propagation over longer
distances; it appears near $T_{\mathrm{quant}}$ and expands in radius only near
$T_{\mathrm{coh}}$, providing a real-space manifestation of the separation
between local quantum dynamics and global quantum coherence that defines the
semi-quantum regime.
Generalizing from these observations, we conjecture
that
local caging
and an approximately $T$-linear inverse
diffusivity
may be
general features of semi-quantum liquids in more
realistic models of strongly interacting electrons (possibly including interactions with phonons), even though the
microscopic nature of the cages
surely differs from case to case.

\section{Semi-quantum bad-metal transport}
\label{sec:transport}

\begin{figure}[t]
    \centering
\includegraphics[width=\textwidth]{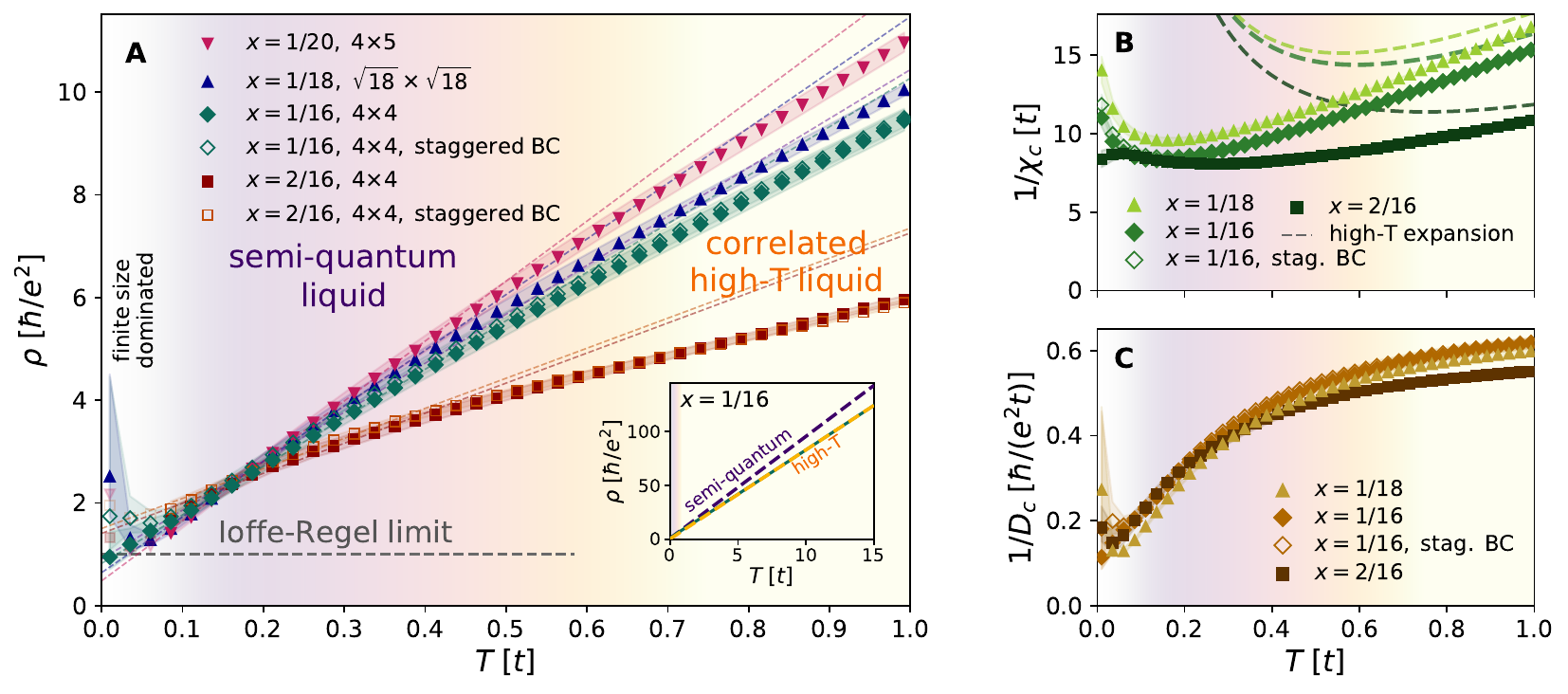}
    \caption{\textbf{Bad-metallic transport
    regimes.}
(A) Twist-averaged dc resistivity for $x=1/20$, $1/18$, $1/16$, and $2/16$. The resistivity
is approximately $T$-linear and above the Ioffe--Regel scale, but with a slope change near $T_{\rm quant}$:
the dashed lines  are linear fits to the data in the semi-quantum regime.
The inset compares linear fits to $\rho(T)$ in the high-$T$ and semi-quantum regimes over an expanded range of $T$.
Results are shown for non-staggered and staggered boundary conditions.
(B,C) Nernst--Einstein decomposition, $\rho=1/(\chi_cD_c)$.
The dashed curve in (B) is the first-order high-temperature expansion. Shaded bands indicate the standard
deviation over different twisted boundary conditions.
}
    \label{fig:rho_kappa_D}
\end{figure}

Figure~\ref{fig:rho_kappa_D} presents the central result. The dc resistivity is approximately $T$-linear and lies above the Ioffe--Regel limit across the
temperature window  $T_{\rm coh}\lesssim T\leq t$ (i.e. ending somewhere in the crossover regime between the semi-quantum and high temperature behavior) for dopings $x=1/20$, $1/18$, $1/16$, and $2/16$ (Fig.~\ref{fig:rho_kappa_D}A). We take the Ioffe--Regel scale of a two-dimensional metal to be $\rho_{\mathrm{IR}}=\hbar/e^{2}$, corresponding to $k_F\ell\sim2\pi$ (horizontal dashed line in Fig.~\ref{fig:rho_kappa_D}A). Dashed lines in Fig.~\ref{fig:rho_kappa_D}A
are $T$-linear fits to $\rho$ in the semi-quantum regime;
the inset highlights the distinct $T$-linear slopes in the high-$T$ and semi-quantum regimes, shown over a wider range of temperatures.
The $T$-linearity in the two regimes
have  physically distinct origins, revealed by the Nernst--Einstein relation $\rho=1/(\chi_c D_c)$, which expresses the resistivity as the product of a thermodynamic factor, the compressibility $\chi_c$, and a dynamical factor, the charge diffusivity $D_c$ (Fig.~\ref{fig:rho_kappa_D}B,C). At high temperatures, the inverse compressibility $1/\chi_c$ rises approximately linearly with $T$, as expected from the high-temperature expansion (dashed curve, Fig.~\ref{fig:rho_kappa_D}B; see also App.~\ref{sec:HTE}); in this regime the $T$-linear resistivity is largely thermodynamic in origin. Below $T_{\rm quant}\approx0.5$--$0.8t$, identified operationally by the departure of $1/\chi_c$ from the high-temperature form, the inverse compressibility saturates toward a nearly constant value, signaling the
importance of quantum dynamics. The dominant temperature dependence of $\rho$ has shifted from thermodynamics to dynamics: it is now carried by the inverse diffusivity $1/D_c$, which becomes approximately $T$-linear in the semi-quantum regime (Fig.~\ref{fig:rho_kappa_D}C).
This is the key observation: Below $T_{\rm quant}$, bad metallicity is no longer the high-temperature thermodynamic contribution extended downward; instead it is a property of incoherent charge quantum dynamics, encoded in the diffusivity $D_c$.

The explicit calculations we have carried out are on clusters with toroidal geometries. As one way of assessing finite-size effects, we compare results for different periodic cluster connectivities (distinct Betts cluster geometries \cite{Betts1999}), which we refer to as ``staggered'' and ``non-staggered'' and illustrate in Fig.~\ref{fig:geometries} of App.~\ref{sec:model}.
Moreover, to reduce finite-size effects, for each cluster geometry,  we apply boundary conditions corresponding to phase ``twists''  $\Delta \theta_a,\, a=x,y$, in the two orthogonal directions around the torus (which can be thought of as corresponding to fractions of a flux quantum through the holes of the torus), where unless otherwise stated,
calculations are carried out for 25 values of $\Delta \theta_a = 2\pi\left(n_a+ \varphi\right)/5$
with $0 \leq n_a < 5$ and an irrational offset $\varphi = (1+\sqrt5)/2$.
As is illustrated in Fig.~\ref{fig:rho_kappa_D},
the differences in the results for the staggered and unstaggered geometries
and the twist variance of the results are
small for $T\gtrsim0.08t$.
This weak
sensitivity to
cluster geometry is also consistent with transport being
controlled primarily by short-range correlations.
At lower temperatures, the twist variance grows and the two
geometric sectors
separate, marking the onset of a finite-size-dominated regime
at $T$ near
$T_{\rm coh}$. Fortunately, the accessible $T$ window in which finite-size effects are plausibly negligible
includes
the relevant
semi-quantum regime.

\section{Optical and thermodynamic signatures}
\label{sec:optical_thermo}

\begin{figure}[t]
    \centering
\includegraphics[width=1\textwidth]{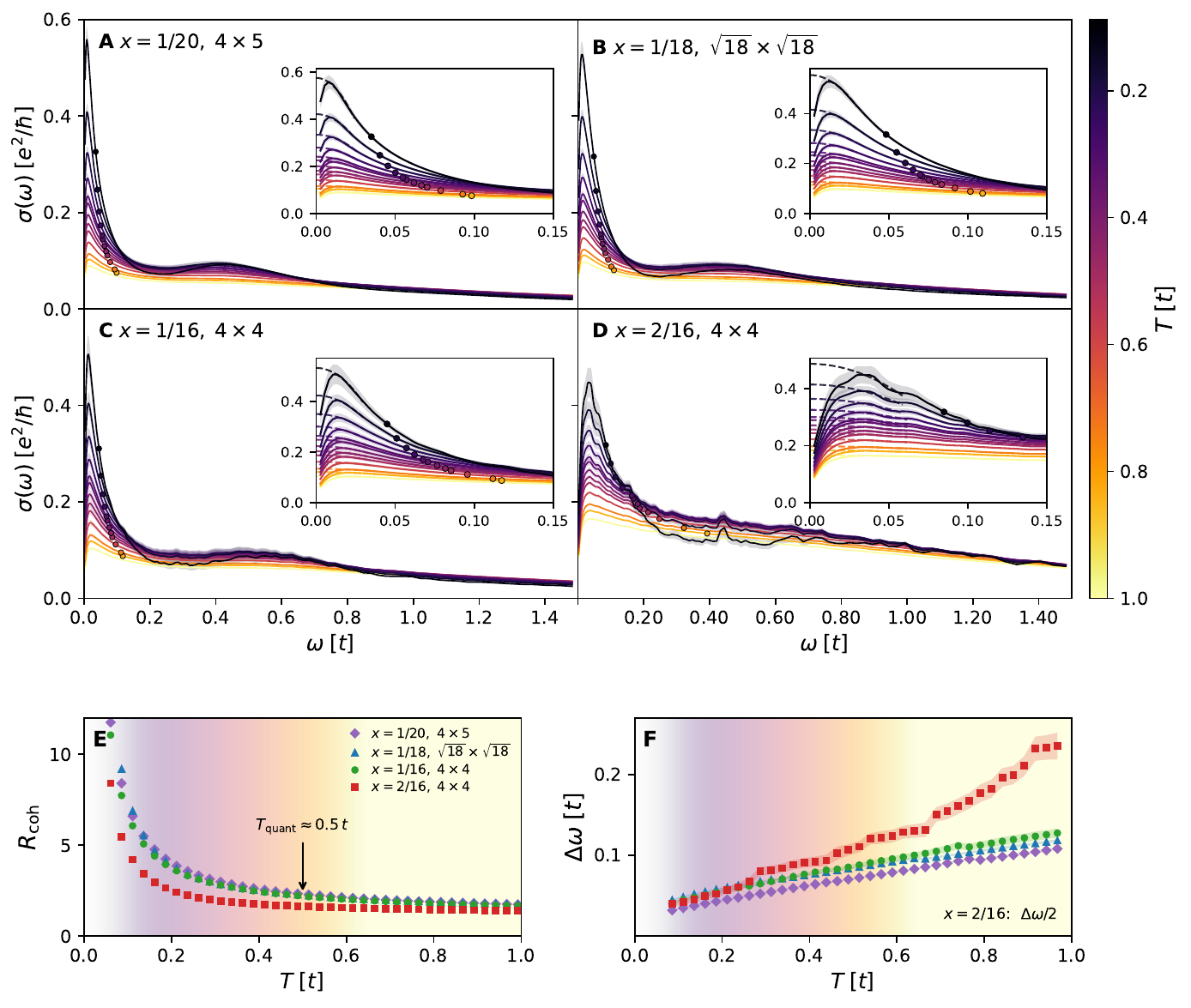}
\caption{\textbf{Optical conductivity
in the semi-quantum
regime.}
({\bf A--D}) optical conductivity $\sigma(\omega,T)$ for $x=1/20$ ({\bf A}; $4\times5$), $x=1/18$
({\bf B}; $\sqrt{18}\times\sqrt{18}$), and $x=1/16$, $2/16$ ({\bf C, D}; $4\times4$),
with curves colored by temperature. Shaded bands indicate the standard error
over boundary twists; insets show the low-frequency extrapolation used to
obtain $\sigma_{\rm dc}$.
The low frequency portion of $\sigma$ can be  characterized  by ({\bf E}) a weight, $R_{\rm coh}=\sigma_{\rm dc}/\sigma(\omega=0.5t)$, and ({\bf F}) a width, $\Delta\omega(T)$, defined by $\Delta\sigma(\Delta\omega,T)=\tfrac{1}{2}
\Delta\sigma_{\rm dc}(T)$, with $\Delta\omega>\omega_{\max}$,
$\omega_{\max}$ being the position of the maximum of
$\Delta\sigma(\omega,T)$.
The $x=2/16$ values in ({\bf F}) are multiplied by $1/2$ for visual
comparison.  (Values of $\sigma$ at $\Delta\omega(T)$ are also indicated by the filled circles in A--D.)
}
    \label{fig:optical conductivity}
\end{figure}

The optical conductivity reveals how charge dynamics evolve across the crossover
into the semi-quantum regime (Fig.~\ref{fig:optical conductivity}, A--D;
shaded bands show the twist variance).
At high temperatures,
$\sigma(\omega,T)$ is broad and nearly featureless, consistent with incoherent
charge motion. Upon cooling below $T_{\rm quant}$, low-frequency spectral weight
builds up into an enhanced peak at low frequencies. The response, however, remains
broad -- on the scale of $t$; it is an incipient Drude-like feature but not a sharp
quasiparticle peak.
Over the same temperature range, the one-hole optical spectra also develop a broad
finite-frequency feature at frequencies of order $t$, rather than a clean
separation into a narrow Drude peak and a high-energy background.

The low-frequency insets show a slight downturn as $\omega\to0$, which we
attribute to finite-cluster artifacts~\cite{MukerjeeOganesyanHuse2006}. We obtain
$\sigma_{\mathrm{dc}}$ from the quadratic low-frequency extrapolation
illustrated in the insets; varying the fitting window and functional form
changes $\sigma_{\mathrm{dc}}$ by less than a few percent for
$T\gtrsim 0.08t$, comparable to the twist standard deviation.
Details are given in App.~\ref{sec:dcextraction}.

We quantify
the buildup and
narrowing of the
low-frequency response as follows:
We define a ``coherence ratio,''
$R_{\rm coh}(T)=\sigma_{\rm dc}(T)/\sigma(\omega=0.5t,T)$.
As shown in Fig.~\ref{fig:optical conductivity}E, $R_{\rm coh}$ begins to grow
near $T_{\rm quant}$, providing an optical signature of the crossover into the
semi-quantum regime. A sharply coherent Drude response would give
$R_{\rm coh}\gg1$, whereas throughout most of the semi-quantum regime it remains
$\mathcal{O}(1)$, consistent with a broad low-frequency response rather than a
coherent Drude peak.
To characterize the narrowing of the low-frequency feature without assuming a
specific functional form, we define the change in optical conductivity relative
to a fixed high-temperature reference,
$\Delta\sigma(\omega,T)=\sigma(\omega,T)-\sigma(\omega,T_{\rm ref})$,
with $T_{\rm ref}=t$. We characterize the frequency extent of the
low-frequency feature by its high-frequency half-maximum edge,
$\Delta\omega(T)$, defined by
$\Delta\sigma(\Delta\omega,T)=\tfrac{1}{2}
\Delta\sigma_{\rm dc}(T)$, with $\Delta\omega>\omega_{\max}$, where
$\omega_{\max}$ denotes the position of the maximum of
$\Delta\sigma(\omega,T)$.
As shown in
Fig.~\ref{fig:optical conductivity}F, $\Delta\omega$ decreases approximately
linearly upon cooling for the one-hole clusters, demonstrating the progressive
narrowing of the low-frequency feature.
(The filled circles in (A--D) mark $\sigma(\omega,T)$ at $\omega=\Delta\omega(T)$.)
Note that the two-hole results
are qualitatively
distinct, with a larger characteristic frequency scale and a different
temperature dependence. With the system sizes presently accessible, we cannot
determine whether these differences reflect the higher doping or finite-size
effects. For visual comparison, the $x=2/16$ values in
Fig.~\ref{fig:optical conductivity}F are multiplied by $1/2$. Changing the
reference temperature shifts the absolute values of $\Delta \omega$ somewhat, particularly at low
temperature,
but leaves the qualitative temperature dependence
unchanged (see App.~\ref{sec:lineshape}).

\begin{figure}[t]
    \centering
\includegraphics[width=\textwidth]{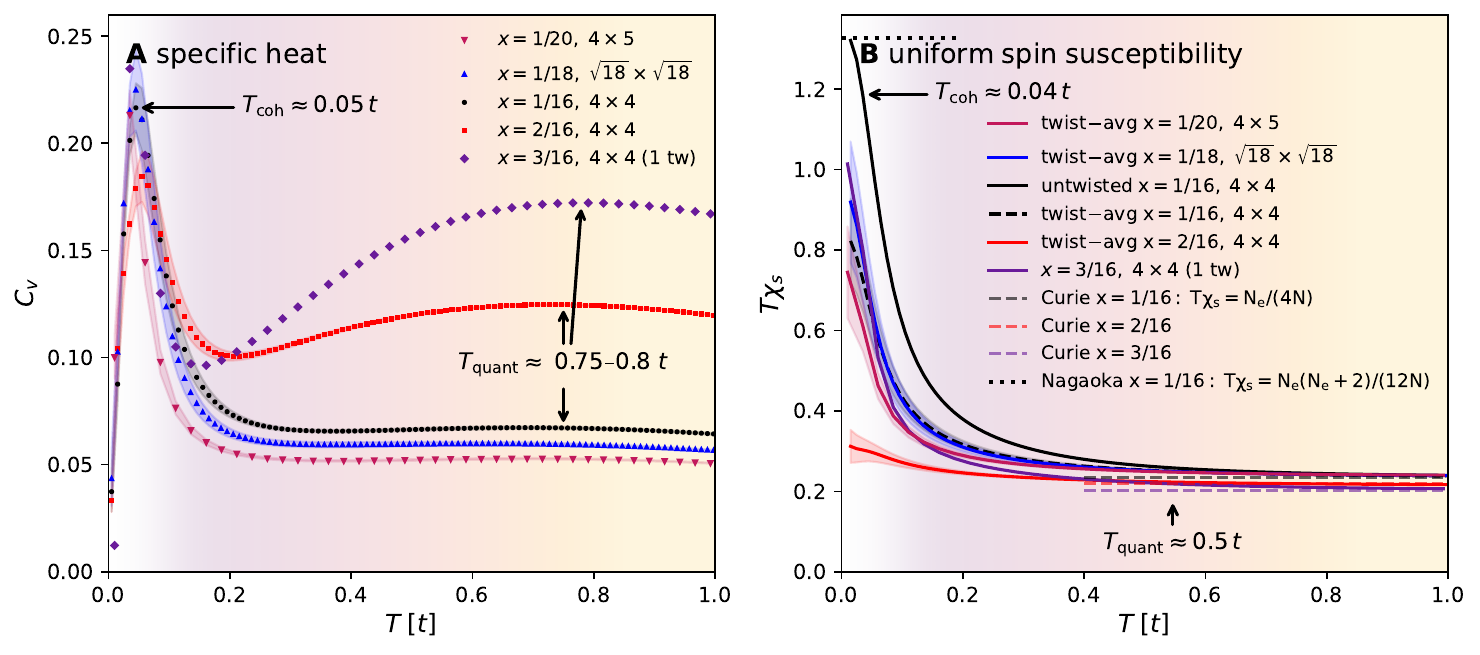}
    \caption{\textbf{Thermodynamics across the semi-quantum crossover.}
({\bf A}) Specific heat per site for $x=1/20,1/18$, $1/16$, $2/16$, and $3/16$.
({\bf B}) Uniform spin susceptibility for the indicated dopings and boundary
conditions; dashed lines show the Curie and finite-cluster Nagaoka values.
Shaded bands indicate the standard error over boundary twists. Data for
$x=3/16$ are shown for a single irrational twist.
}
    \label{fig:thermo}
\end{figure}

The specific heat and uniform spin susceptibility provide complementary thermodynamic signatures of the semi-quantum crossover (Fig.~\ref{fig:thermo}). In panel A, $C_V$ develops a broad maximum near $T_{\rm quant}\approx0.75t$, marking its departure from the classical high-temperature behavior $C_V\sim1/T^2$, followed by a much sharper peak near the lower scale $T_{\rm coh}$ associated with the approach to the spin-polarized ferromagnetic ground state. Panel B shows $\chi_s=(\langle S_z^2\rangle-\langle S_z\rangle^2)/(NT)$ for $x=1/18$, $1/16$, and $2/16$. All dopings depart near $T_{\rm quant}$ from the independent-moment Curie form, $\chi_s=(1-x)/(4T)$. For the untwisted single-hole $4\times4$ cluster, $\chi_s$ approaches the finite-cluster Nagaoka value, $\chi_s^{\rm Nag}=N_e(N_e+2)/(12NT)$, only at much lower temperature \cite{RieraYoung1989}. We define $T_{\rm coh}$ operationally as the temperature at which the untwisted susceptibility lies within $10\%$ of $\chi_s^{\rm Nag}$, consistent with $T_{\rm coh}\sim xt$. Together with the charge compressibility, the specific heat and spin susceptibility show that the semi-quantum crossover occurs near the same scale across the charge, spin, and energy sectors.

\section{Formation of a spin cage}
\label{sec:cage}

The Andreev--Kosevich picture suggests a concrete real-space criterion for the electronic semi-quantum regime. At $T_{\mathrm{quant}}$, the onset of quantum dynamics should be
accompanied with the existence of a finite local
``cage'' in which  there are well-separated discrete quantum states of an individual carrier,
but this environment should remain spatially bounded until coherent motion emerges near $T_{\mathrm{coh}}$.
In the infinite-$U$ Hubbard model,
the cage is provided not by positional correlations but by the fluctuating spin background through which the doped hole moves. Successive hole hops permute the spins along its path, so motion through a disordered background continually rearranges the spin configuration \cite{BrinkmanRice1970}. Local ferromagnetic correlations reduce
the  kinetic frustration, allowing the hole to move more effectively within a polarized region. At finite temperature, however, the kinetic-energy gain competes with the entropic cost of polarizing an extended region. The result
is a finite ferromagnetic environment within which hole motion is
coherent, but beyond which it again encounters a disordered and kinetically frustrating spin background. It is in this sense that the polarized region constitutes a
cage.

We probe this structure directly using the connected hole--spin-bond
correlator
\begin{equation}
C_h(\mathbf R,\mathbf e;T)
=
\left\langle
\rho_h(\mathbf 0)\,
\mathbf S_{\mathbf R-\mathbf e/2}\!\cdot\!
\mathbf S_{\mathbf R+\mathbf e/2}
\right\rangle_T
-
\left\langle\rho_h(\mathbf 0)\right\rangle_T
\left\langle
\mathbf S_{\mathbf R-\mathbf e/2}\!\cdot\!
\mathbf S_{\mathbf R+\mathbf e/2}
\right\rangle_T .
\label{eq:nSS}
\end{equation}
Here, $\rho_h(\mathbf 0)$ is the hole density at the origin, $\mathbf R$ is
the displacement from the hole to the center of a nearest-neighbor bond
oriented in the
$\mathbf e=\hat{\mathbf x},\hat{\mathbf y}$
direction.
Thus, $C_h$
is the change in the local spin correlations
induced by
the presence of
a nearby hole.

\begin{figure}[t]
    \centering
    \includegraphics[width=0.6\textwidth]{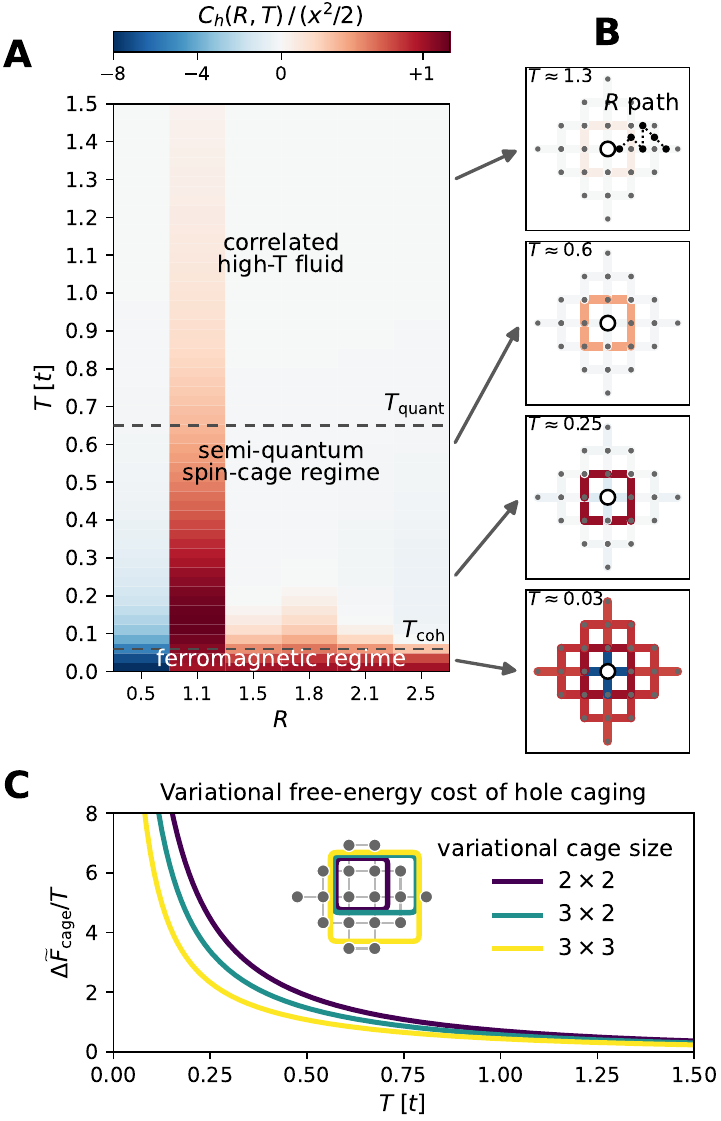}
    \caption{\textbf{Formation and variational cost of a spin cage.}
\textbf{A}, Connected hole--spin-bond correlator
$C_h(R,T)$ for a representative boundary twist on the 18-site cluster, normalized by $x^2/2$, the value of $C_h$ in the Nagaoka ground state.
\textbf{B}, Corresponding bond-resolved patterns of
$C_h(\mathbf R,\mathbf e;T)$ at selected temperatures; the path used to
define $R$ in panel ({\bf A}) is indicated in the top panel of ({\bf B}).
\textbf{C}, Variational free-energy cost $\Delta \widetilde{F}_{\rm cage}/T$ of
restricting the hole to the indicated finite regions as a function of $T$.
}
    \label{fig:cage}
\end{figure}

Figure~\ref{fig:cage}A shows the radial profile $C_h(R,T)$ of the hole-induced
spin correlations, extracted from $C_h(\mathbf R,\mathbf e;T)$ using bond centers that lie a distance $R=|\mathbf R|$ from the hole, normalized by the value
in the Nagaoka ferromagnetic ground state $x^2/2$ (App.~\ref{sec:triplecorr}). The results
are evaluated for an untwisted boundary on the 18-site cluster;
Fig.~\ref{fig:cage}B shows the corresponding bond-resolved textures. As $T$ is lowered from above to below $T_{\rm quant}$, the hole gradually polarizes its immediate
surroundings, while the more distant spin background remains weakly affected.
The resulting finite-radius ferromagnetic texture defines the \emph{spin cage}.
Only upon cooling below $T_{\rm coh}$ does the ferromagnetic correlation length grow beyond the
cluster size, marking the crossover to coherent (Fermi liquid) charge motion and the approach
to the $T=0$ ferromagnetic ground state.

The formation of a finite polarized region around the hole, embedded in an
otherwise fluctuating spin background, suggests that the hole motion may be
effectively restricted to this spin cage. To test this, we construct a trial
system in which the hole is confined to a subregion comparable in size to the
spin texture while the exterior remains half filled, and compute the
corresponding free energy $F_{\rm tr}$, which provides a variational upper
bound on the full free energy $F$ (App.~\ref{sec:caging}). At high
temperatures, $\Delta F_{\rm cage}\equiv F_{\rm tr}-F$ approaches
$T\ln(N/N_{\rm cage})$, the free-energy cost associated with the loss of
positional entropy upon restricting the hole to $N_{\rm cage}$ of the $N$
sites. We therefore consider
$\Delta\widetilde F_{\rm cage}\equiv\Delta F_{\rm cage}-T\ln(N/N_{\rm cage})$,
the confinement cost beyond this positional contribution. As shown in
Fig.~\ref{fig:cage}C, at high temperatures this residual cost is small,
consistent with the approximate locality of the free energy. Around
$T_{\rm quant}$ we find $\Delta\widetilde F_{\rm cage}\sim T$, and
$\Delta\widetilde F_{\rm cage}/T$ remains of order unity through much of the
semi-quantum regime. Thus, although the free energy is no longer strictly
local, confinement on the scale of the spin texture costs of order the
thermal scale, consistent with a fluctuating rather than static cage. On
approaching $T_{\rm coh}$, $\Delta\widetilde F_{\rm cage}/T$ grows rapidly,
signaling the breakdown of this local confinement description as
ferromagnetic correlations spread across the system.

\section{Discussion and outlook}
\label{sec:discussion}

Our discussion thus far has largely treated the calculations as a numerical experiment, focusing on the essential features of the semi-quantum regime with minimal interpretation. Most observables are consistent with a dilute, yet strongly correlated limit, with relatively small subleading corrections in $x$. The two-hole transport results show larger deviations, including a broader low-frequency optical response, although with the system sizes presently accessible we cannot distinguish genuine doping dependence from finite-size effects. A more puzzling feature revealed by our results is the behavior of the compressibility in the semi-quantum regime: it becomes approximately independent of both $T$ and $x$, yet remains about $1.5$ times larger than the $T=0$ Nagaoka value of the spin-polarized Fermi liquid, $\chi_c=1/(4\pi t)$, approaching that value only at much lower temperatures near $T_{\rm coh}$. Thus, the nearly $T$-independent quantum response that develops below $T_{\rm quant}$ remains distinct from the eventual low-temperature quasiparticle regime. A natural starting point for a theory of the semi-quantum regime may be the dynamics of a single caged hole, with some similarity to local descriptions underlying DMFT studies of bad metallicity. However, as shown in App.~\ref{sec:bubble_cond}, vertex corrections must be retained for quantitative transport: comparing the full Kubo result with the bubble conductivity constructed from the same exact spectral function isolates a sizable vertex contribution,
as also suggested by related comparisons of the finite-$U$ Hubbard model~\cite{Brown2019,VucicevicKokaljZitko2019Vertex}. The short-ranged correlations nevertheless suggest that a tractable quasi-local description may be possible.

We now comment on what we see as the physical relevance of our results to the properties of  strongly correlated metals.
The central significance of our results lies neither in the conventional low-temperature coherent regime nor in the asymptotic high-temperature limit, but in the intermediate regime separating them.
The low $T$ behavior has much in common with behavior of many materials in that it can be understood in terms of broken symmetries and well-defined quasi-particles.  The high $T$ behavior is probably not relevant to typical quantum materials, since it derives from the boundedness of the spectrum of the tight -binding model - although it may be relevant in certain ``flat band'' contexts.  Our key finding  is that - rather than a smooth crossover between these two extremal regimes - there is a well-defined intermediate semi-quantum regime in which the particles are largely confined in relatively small emergent cages.
The microscopic origin of the \(T\)-linear inverse diffusivity remains to be understood. Nevertheless,
the resistivity in this regime is roughly $T-$linear, as in the high $T$ regime but with a different slope and extrapolated intercept, while thermodynamic quantities, such as the compressibility, exhibit a $T$ dependence similar to that in the low $T$ regime, albeit with different magnitudes.
The physics driving this is local - and indeed it likely does not depend in any significant way on the quantum statistics of the particles.

Cold-atom systems are particularly promising for testing this picture because local correlations can in principle be imaged directly and related to transport in the same controlled setting. Local dopant--spin correlations closely related to the spin cages identified here have already been resolved in Fermi--Hubbard quantum simulators \cite{Koepsell2019,Ji2021MagneticPolaron,Koepsell2021,Lebrat2024,Prichard2024}, while bad-metal transport and charge diffusion have been measured in cold-atom Hubbard systems \cite{Brown2019,Xu2019}. Such platforms therefore offer the prospect of tracking the temperature evolution of local caging alongside charge diffusion, directly testing their interplay in the semi-quantum regime. Moreover, if the essential physics is independent of particle statistics, analogous behavior may also be accessible in strongly interacting bosonic
systems, broadening the range of platforms in which this physics can be explored.

We conclude by speculating that the bad-metal regimes of a broader class of correlated electronic systems may reflect analogous semi-quantum physics.
The nature of the ``cages'' will naturally vary greatly from case to case - it can be local nearly crystalline environments as in liquid He (and possibly in the 2D electron fluid at large $r_s$ \cite{Spivak2010}) or a polaronic distortion in the case of strong electron-phonon coupling, or can even be largely structural in character, as for example in a C$_{60}$ molecule in A$_3$C$_{60}$ \cite{Hebard1993,Gunnarsson1997}.
It is at least intuitively clear how such local fluctuating cage structures can lead to long incoherent tails in the optical conductivity.
What remains less clear is why, and under what general conditions, they produce both the buildup of low-frequency spectral weight and \(T\)-linear resistivity. The present results suggest that a theory of this kind may be possible using only the physics of strong local correlations.


\begin{acknowledgments}
SAK particularly thanks B. Spivak for his obsession with the semiquantum fluid over many years.  We also acknowledge significant discussions with Srinivas Raghu, Noga Bashan, Erez Berg, Pavel Nosov, Antoine Georges, Andrey Chubukov and Lev Ioffe. ET thanks Guy Tevet for a useful discussion on GPU-based diagonalization. ChatGPT (OpenAI) and Claude (Anthropic) were used as technical aids for code optimization and debugging, preparation of plotting scripts, and manuscript editing. All scientific methodology, numerical results, analysis, and interpretation were developed and validated by the authors.
ET was supported in part by NSF-BSF award DMR-2310312, a research gift from Periodic Labs, Inc.,the Zuckerman STEM fellowship and the Geballe
Laboratory of Advanced Materials Postdoctoral Fellowship, TPD and SAK were supported
by  the U.S. Department of Energy (DOE), Office of Science, Basic Energy Sciences, Materials Sciences and Engineering Division, under contract DE-AC02-76SF00515. Computational work was performed on the Sherlock cluster at Stanford University and
on resources of the National Energy Research Scientific
Computing Center, supported by the U.S. Department
of Energy under award BES-ERCAP0034969.
\end{acknowledgments}


\bibliography{refs.bib}

\begin{thebibliography}{56}%
\makeatletter
\providecommand \@ifxundefined [1]{%
 \@ifx{#1\undefined}
}%
\providecommand \@ifnum [1]{%
 \ifnum #1\expandafter \@firstoftwo
 \else \expandafter \@secondoftwo
 \fi
}%
\providecommand \@ifx [1]{%
 \ifx #1\expandafter \@firstoftwo
 \else \expandafter \@secondoftwo
 \fi
}%
\providecommand \natexlab [1]{#1}%
\providecommand \enquote  [1]{``#1''}%
\providecommand \bibnamefont  [1]{#1}%
\providecommand \bibfnamefont [1]{#1}%
\providecommand \citenamefont [1]{#1}%
\providecommand \href@noop [0]{\@secondoftwo}%
\providecommand \href [0]{\begingroup \@sanitize@url \@href}%
\providecommand \@href[1]{\@@startlink{#1}\@@href}%
\providecommand \@@href[1]{\endgroup#1\@@endlink}%
\providecommand \@sanitize@url [0]{\catcode `\\12\catcode `\$12\catcode `\&12\catcode `\#12\catcode `\^12\catcode `\_12\catcode `\%12\relax}%
\providecommand \@@startlink[1]{}%
\providecommand \@@endlink[0]{}%
\providecommand \url  [0]{\begingroup\@sanitize@url \@url }%
\providecommand \@url [1]{\endgroup\@href {#1}{\urlprefix }}%
\providecommand \urlprefix  [0]{URL }%
\providecommand \Eprint [0]{\href }%
\providecommand \doibase [0]{https://doi.org/}%
\providecommand \selectlanguage [0]{\@gobble}%
\providecommand \bibinfo  [0]{\@secondoftwo}%
\providecommand \bibfield  [0]{\@secondoftwo}%
\providecommand \translation [1]{[#1]}%
\providecommand \BibitemOpen [0]{}%
\providecommand \bibitemStop [0]{}%
\providecommand \bibitemNoStop [0]{.\EOS\space}%
\providecommand \EOS [0]{\spacefactor3000\relax}%
\providecommand \BibitemShut  [1]{\csname bibitem#1\endcsname}%
\let\auto@bib@innerbib\@empty
\bibitem [{\citenamefont {Ioffe}\ and\ \citenamefont {Regel}(1960)}]{IoffeRegel1960}%
  \BibitemOpen
  \bibfield  {author} {\bibinfo {author} {\bibfnamefont {A.~F.}\ \bibnamefont {Ioffe}}\ and\ \bibinfo {author} {\bibfnamefont {A.~R.}\ \bibnamefont {Regel}},\ }\bibfield  {title} {\bibinfo {title} {Non-crystalline, amorphous and liquid electronic semiconductors},\ }\href@noop {} {\bibfield  {journal} {\bibinfo  {journal} {Progress in Semiconductors}\ }\textbf {\bibinfo {volume} {4}},\ \bibinfo {pages} {237} (\bibinfo {year} {1960})}\BibitemShut {NoStop}%
\bibitem [{\citenamefont {Gunnarsson}\ \emph {et~al.}(2003)\citenamefont {Gunnarsson}, \citenamefont {Calandra},\ and\ \citenamefont {Han}}]{Gunnarsson2003}%
  \BibitemOpen
  \bibfield  {author} {\bibinfo {author} {\bibfnamefont {O.}~\bibnamefont {Gunnarsson}}, \bibinfo {author} {\bibfnamefont {M.}~\bibnamefont {Calandra}},\ and\ \bibinfo {author} {\bibfnamefont {J.~E.}\ \bibnamefont {Han}},\ }\bibfield  {title} {\bibinfo {title} {Colloquium: Saturation of electrical resistivity},\ }\href {https://doi.org/10.1103/RevModPhys.75.1085} {\bibfield  {journal} {\bibinfo  {journal} {Reviews of Modern Physics}\ }\textbf {\bibinfo {volume} {75}},\ \bibinfo {pages} {1085} (\bibinfo {year} {2003})}\BibitemShut {NoStop}%
\bibitem [{\citenamefont {Emery}\ and\ \citenamefont {Kivelson}(1995)}]{EmeryKivelson1995}%
  \BibitemOpen
  \bibfield  {author} {\bibinfo {author} {\bibfnamefont {V.~J.}\ \bibnamefont {Emery}}\ and\ \bibinfo {author} {\bibfnamefont {S.~A.}\ \bibnamefont {Kivelson}},\ }\bibfield  {title} {\bibinfo {title} {Superconductivity in bad metals},\ }\href {https://doi.org/10.1103/PhysRevLett.74.3253} {\bibfield  {journal} {\bibinfo  {journal} {Physical Review Letters}\ }\textbf {\bibinfo {volume} {74}},\ \bibinfo {pages} {3253} (\bibinfo {year} {1995})}\BibitemShut {NoStop}%
\bibitem [{\citenamefont {Hussey}\ \emph {et~al.}(2004)\citenamefont {Hussey}, \citenamefont {Takenaka},\ and\ \citenamefont {Takagi}}]{Hussey2004}%
  \BibitemOpen
  \bibfield  {author} {\bibinfo {author} {\bibfnamefont {N.~E.}\ \bibnamefont {Hussey}}, \bibinfo {author} {\bibfnamefont {K.}~\bibnamefont {Takenaka}},\ and\ \bibinfo {author} {\bibfnamefont {H.}~\bibnamefont {Takagi}},\ }\bibfield  {title} {\bibinfo {title} {Universality of the mott--ioffe--regel limit in metals},\ }\href {https://doi.org/10.1080/14786430410001716944} {\bibfield  {journal} {\bibinfo  {journal} {Philosophical Magazine}\ }\textbf {\bibinfo {volume} {84}},\ \bibinfo {pages} {2847} (\bibinfo {year} {2004})}\BibitemShut {NoStop}%
\bibitem [{\citenamefont {Phillips}\ \emph {et~al.}(2022)\citenamefont {Phillips}, \citenamefont {Hussey},\ and\ \citenamefont {Abbamonte}}]{Phillips2022}%
  \BibitemOpen
  \bibfield  {author} {\bibinfo {author} {\bibfnamefont {P.~W.}\ \bibnamefont {Phillips}}, \bibinfo {author} {\bibfnamefont {N.~E.}\ \bibnamefont {Hussey}},\ and\ \bibinfo {author} {\bibfnamefont {P.}~\bibnamefont {Abbamonte}},\ }\bibfield  {title} {\bibinfo {title} {Stranger than metals},\ }\href {https://doi.org/10.1126/science.abh4273} {\bibfield  {journal} {\bibinfo  {journal} {Science}\ }\textbf {\bibinfo {volume} {377}},\ \bibinfo {pages} {eabh4273} (\bibinfo {year} {2022})}\BibitemShut {NoStop}%
\bibitem [{\citenamefont {Keimer}\ \emph {et~al.}(2015)\citenamefont {Keimer}, \citenamefont {Kivelson}, \citenamefont {Norman}, \citenamefont {Uchida},\ and\ \citenamefont {Zaanen}}]{Keimer2015}%
  \BibitemOpen
  \bibfield  {author} {\bibinfo {author} {\bibfnamefont {B.}~\bibnamefont {Keimer}}, \bibinfo {author} {\bibfnamefont {S.~A.}\ \bibnamefont {Kivelson}}, \bibinfo {author} {\bibfnamefont {M.~R.}\ \bibnamefont {Norman}}, \bibinfo {author} {\bibfnamefont {S.}~\bibnamefont {Uchida}},\ and\ \bibinfo {author} {\bibfnamefont {J.}~\bibnamefont {Zaanen}},\ }\bibfield  {title} {\bibinfo {title} {From quantum matter to high-temperature superconductivity in copper oxides},\ }\href {https://doi.org/10.1038/nature14165} {\bibfield  {journal} {\bibinfo  {journal} {Nature}\ }\textbf {\bibinfo {volume} {518}},\ \bibinfo {pages} {179} (\bibinfo {year} {2015})}\BibitemShut {NoStop}%
\bibitem [{\citenamefont {Lee}\ \emph {et~al.}(2023)\citenamefont {Lee}, \citenamefont {Wang}, \citenamefont {Osada}, \citenamefont {Goodge}, \citenamefont {Wang}, \citenamefont {Lee}, \citenamefont {Harvey}, \citenamefont {Kim}, \citenamefont {Yu}, \citenamefont {Murthy}, \citenamefont {Raghu}, \citenamefont {Kourkoutis},\ and\ \citenamefont {Hwang}}]{Lee2023}%
  \BibitemOpen
  \bibfield  {author} {\bibinfo {author} {\bibfnamefont {K.}~\bibnamefont {Lee}}, \bibinfo {author} {\bibfnamefont {B.~Y.}\ \bibnamefont {Wang}}, \bibinfo {author} {\bibfnamefont {M.}~\bibnamefont {Osada}}, \bibinfo {author} {\bibfnamefont {B.~H.}\ \bibnamefont {Goodge}}, \bibinfo {author} {\bibfnamefont {T.~C.}\ \bibnamefont {Wang}}, \bibinfo {author} {\bibfnamefont {Y.}~\bibnamefont {Lee}}, \bibinfo {author} {\bibfnamefont {S.~P.}\ \bibnamefont {Harvey}}, \bibinfo {author} {\bibfnamefont {W.~J.}\ \bibnamefont {Kim}}, \bibinfo {author} {\bibfnamefont {Y.}~\bibnamefont {Yu}}, \bibinfo {author} {\bibfnamefont {C.}~\bibnamefont {Murthy}}, \bibinfo {author} {\bibfnamefont {S.}~\bibnamefont {Raghu}}, \bibinfo {author} {\bibfnamefont {L.~F.}\ \bibnamefont {Kourkoutis}},\ and\ \bibinfo {author} {\bibfnamefont {H.~Y.}\ \bibnamefont {Hwang}},\ }\bibfield  {title} {\bibinfo {title} {Linear-in-temperature resistivity for optimally superconducting {(Nd,Sr)NiO$_2$}},\ }\href {https://doi.org/10.1038/s41586-023-06129-x}
  {\bibfield  {journal} {\bibinfo  {journal} {Nature}\ }\textbf {\bibinfo {volume} {619}},\ \bibinfo {pages} {288} (\bibinfo {year} {2023})}\BibitemShut {NoStop}%
\bibitem [{\citenamefont {Hsu}\ \emph {et~al.}(2024)\citenamefont {Hsu}, \citenamefont {Lee}, \citenamefont {Badoux}, \citenamefont {Duffy}, \citenamefont {Cuoghi}, \citenamefont {Wang}, \citenamefont {Kool}, \citenamefont {Ha{\"i}k-Dunn}, \citenamefont {Hwang},\ and\ \citenamefont {Hussey}}]{Hsu2024}%
  \BibitemOpen
  \bibfield  {author} {\bibinfo {author} {\bibfnamefont {Y.-T.}\ \bibnamefont {Hsu}}, \bibinfo {author} {\bibfnamefont {K.}~\bibnamefont {Lee}}, \bibinfo {author} {\bibfnamefont {S.}~\bibnamefont {Badoux}}, \bibinfo {author} {\bibfnamefont {C.}~\bibnamefont {Duffy}}, \bibinfo {author} {\bibfnamefont {A.}~\bibnamefont {Cuoghi}}, \bibinfo {author} {\bibfnamefont {B.~Y.}\ \bibnamefont {Wang}}, \bibinfo {author} {\bibfnamefont {A.}~\bibnamefont {Kool}}, \bibinfo {author} {\bibfnamefont {I.}~\bibnamefont {Ha{\"i}k-Dunn}}, \bibinfo {author} {\bibfnamefont {H.~Y.}\ \bibnamefont {Hwang}},\ and\ \bibinfo {author} {\bibfnamefont {N.~E.}\ \bibnamefont {Hussey}},\ }\bibfield  {title} {\bibinfo {title} {Transport phase diagram and anomalous metallicity in superconducting infinite-layer nickelates},\ }\href {https://doi.org/10.1038/s41467-024-54135-y} {\bibfield  {journal} {\bibinfo  {journal} {Nature Communications}\ }\textbf {\bibinfo {volume} {15}},\ \bibinfo {pages} {9863} (\bibinfo {year} {2024})}\BibitemShut
  {NoStop}%
\bibitem [{\citenamefont {Kasahara}\ \emph {et~al.}(2010)\citenamefont {Kasahara}, \citenamefont {Shibauchi}, \citenamefont {Hashimoto}, \citenamefont {Ikada}, \citenamefont {Tonegawa}, \citenamefont {Okazaki}, \citenamefont {Shishido}, \citenamefont {Ikeda}, \citenamefont {Takeya}, \citenamefont {Hirata}, \citenamefont {Terashima},\ and\ \citenamefont {Matsuda}}]{Kasahara2010}%
  \BibitemOpen
  \bibfield  {author} {\bibinfo {author} {\bibfnamefont {S.}~\bibnamefont {Kasahara}}, \bibinfo {author} {\bibfnamefont {T.}~\bibnamefont {Shibauchi}}, \bibinfo {author} {\bibfnamefont {K.}~\bibnamefont {Hashimoto}}, \bibinfo {author} {\bibfnamefont {K.}~\bibnamefont {Ikada}}, \bibinfo {author} {\bibfnamefont {S.}~\bibnamefont {Tonegawa}}, \bibinfo {author} {\bibfnamefont {R.}~\bibnamefont {Okazaki}}, \bibinfo {author} {\bibfnamefont {H.}~\bibnamefont {Shishido}}, \bibinfo {author} {\bibfnamefont {H.}~\bibnamefont {Ikeda}}, \bibinfo {author} {\bibfnamefont {H.}~\bibnamefont {Takeya}}, \bibinfo {author} {\bibfnamefont {K.}~\bibnamefont {Hirata}}, \bibinfo {author} {\bibfnamefont {T.}~\bibnamefont {Terashima}},\ and\ \bibinfo {author} {\bibfnamefont {Y.}~\bibnamefont {Matsuda}},\ }\bibfield  {title} {\bibinfo {title} {Evolution from non-fermi- to fermi-liquid transport via isovalent doping in {BaFe$_2$(As$_{1-x}$P$_x$)$_2$} superconductors},\ }\href {https://doi.org/10.1103/PhysRevB.81.184519} {\bibfield
  {journal} {\bibinfo  {journal} {Physical Review B}\ }\textbf {\bibinfo {volume} {81}},\ \bibinfo {pages} {184519} (\bibinfo {year} {2010})}\BibitemShut {NoStop}%
\bibitem [{\citenamefont {Bloch}\ \emph {et~al.}(1974)\citenamefont {Bloch}, \citenamefont {Cowan},\ and\ \citenamefont {Poehler}}]{BlochCowanPoehler1974}%
  \BibitemOpen
  \bibfield  {author} {\bibinfo {author} {\bibfnamefont {A.~N.}\ \bibnamefont {Bloch}}, \bibinfo {author} {\bibfnamefont {D.~O.}\ \bibnamefont {Cowan}},\ and\ \bibinfo {author} {\bibfnamefont {T.~O.}\ \bibnamefont {Poehler}},\ }\bibfield  {title} {\bibinfo {title} {Organic conductors ii: {TTF--TCNQ} and other organic semimetals},\ }in\ \href@noop {} {\emph {\bibinfo {booktitle} {Energy and Charge Transfer in Organic Semiconductors}}},\ \bibinfo {editor} {edited by\ \bibinfo {editor} {\bibfnamefont {K.}~\bibnamefont {Masuda}}\ and\ \bibinfo {editor} {\bibfnamefont {M.}~\bibnamefont {Silver}}}\ (\bibinfo  {publisher} {Plenum Press},\ \bibinfo {address} {New York},\ \bibinfo {year} {1974})\ pp.\ \bibinfo {pages} {167--174}\BibitemShut {NoStop}%
\bibitem [{\citenamefont {Tyler}\ \emph {et~al.}(1998)\citenamefont {Tyler}, \citenamefont {Mackenzie}, \citenamefont {NishiZaki},\ and\ \citenamefont {Maeno}}]{Tyler1998}%
  \BibitemOpen
  \bibfield  {author} {\bibinfo {author} {\bibfnamefont {A.~W.}\ \bibnamefont {Tyler}}, \bibinfo {author} {\bibfnamefont {A.~P.}\ \bibnamefont {Mackenzie}}, \bibinfo {author} {\bibfnamefont {S.}~\bibnamefont {NishiZaki}},\ and\ \bibinfo {author} {\bibfnamefont {Y.}~\bibnamefont {Maeno}},\ }\bibfield  {title} {\bibinfo {title} {High-temperature resistivity of {Sr$_2$RuO$_4$}: Bad metallic transport in a good metal},\ }\href {https://doi.org/10.1103/PhysRevB.58.R10107} {\bibfield  {journal} {\bibinfo  {journal} {Physical Review B}\ }\textbf {\bibinfo {volume} {58}},\ \bibinfo {pages} {R10107} (\bibinfo {year} {1998})}\BibitemShut {NoStop}%
\bibitem [{\citenamefont {Brown}\ \emph {et~al.}(2019)\citenamefont {Brown}, \citenamefont {Mitra}, \citenamefont {Guardado-Sanchez}, \citenamefont {Nourafkan}, \citenamefont {Reymbaut}, \citenamefont {H{\'e}bert}, \citenamefont {Bergeron}, \citenamefont {Tremblay}, \citenamefont {Kokalj}, \citenamefont {Huse}, \citenamefont {Schauss},\ and\ \citenamefont {Bakr}}]{Brown2019}%
  \BibitemOpen
  \bibfield  {author} {\bibinfo {author} {\bibfnamefont {P.~T.}\ \bibnamefont {Brown}}, \bibinfo {author} {\bibfnamefont {D.}~\bibnamefont {Mitra}}, \bibinfo {author} {\bibfnamefont {E.}~\bibnamefont {Guardado-Sanchez}}, \bibinfo {author} {\bibfnamefont {R.}~\bibnamefont {Nourafkan}}, \bibinfo {author} {\bibfnamefont {A.}~\bibnamefont {Reymbaut}}, \bibinfo {author} {\bibfnamefont {C.-D.}\ \bibnamefont {H{\'e}bert}}, \bibinfo {author} {\bibfnamefont {S.}~\bibnamefont {Bergeron}}, \bibinfo {author} {\bibfnamefont {A.-M.~S.}\ \bibnamefont {Tremblay}}, \bibinfo {author} {\bibfnamefont {J.}~\bibnamefont {Kokalj}}, \bibinfo {author} {\bibfnamefont {D.~A.}\ \bibnamefont {Huse}}, \bibinfo {author} {\bibfnamefont {P.}~\bibnamefont {Schauss}},\ and\ \bibinfo {author} {\bibfnamefont {W.~S.}\ \bibnamefont {Bakr}},\ }\bibfield  {title} {\bibinfo {title} {Bad metallic transport in a cold atom {Fermi--Hubbard} system},\ }\href {https://doi.org/10.1126/science.aat4134} {\bibfield  {journal} {\bibinfo  {journal} {Science}\
  }\textbf {\bibinfo {volume} {363}},\ \bibinfo {pages} {379} (\bibinfo {year} {2019})}\BibitemShut {NoStop}%
\bibitem [{\citenamefont {Cao}\ \emph {et~al.}(2020)\citenamefont {Cao}, \citenamefont {Chowdhury}, \citenamefont {Rodan-Legrain}, \citenamefont {Rubies-Bigorda}, \citenamefont {Watanabe}, \citenamefont {Taniguchi}, \citenamefont {Senthil},\ and\ \citenamefont {Jarillo-Herrero}}]{Cao2020TBGStrangeMetal}%
  \BibitemOpen
  \bibfield  {author} {\bibinfo {author} {\bibfnamefont {Y.}~\bibnamefont {Cao}}, \bibinfo {author} {\bibfnamefont {D.}~\bibnamefont {Chowdhury}}, \bibinfo {author} {\bibfnamefont {D.}~\bibnamefont {Rodan-Legrain}}, \bibinfo {author} {\bibfnamefont {O.}~\bibnamefont {Rubies-Bigorda}}, \bibinfo {author} {\bibfnamefont {K.}~\bibnamefont {Watanabe}}, \bibinfo {author} {\bibfnamefont {T.}~\bibnamefont {Taniguchi}}, \bibinfo {author} {\bibfnamefont {T.}~\bibnamefont {Senthil}},\ and\ \bibinfo {author} {\bibfnamefont {P.}~\bibnamefont {Jarillo-Herrero}},\ }\bibfield  {title} {\bibinfo {title} {Strange metal in magic-angle graphene with near {Planckian} dissipation},\ }\href {https://doi.org/10.1103/PhysRevLett.124.076801} {\bibfield  {journal} {\bibinfo  {journal} {Physical Review Letters}\ }\textbf {\bibinfo {volume} {124}},\ \bibinfo {pages} {076801} (\bibinfo {year} {2020})}\BibitemShut {NoStop}%
\bibitem [{\citenamefont {Arovas}\ \emph {et~al.}(2022)\citenamefont {Arovas}, \citenamefont {Berg}, \citenamefont {Kivelson},\ and\ \citenamefont {Raghu}}]{ArovasBergKivelsonRaghu2022}%
  \BibitemOpen
  \bibfield  {author} {\bibinfo {author} {\bibfnamefont {D.~P.}\ \bibnamefont {Arovas}}, \bibinfo {author} {\bibfnamefont {E.}~\bibnamefont {Berg}}, \bibinfo {author} {\bibfnamefont {S.~A.}\ \bibnamefont {Kivelson}},\ and\ \bibinfo {author} {\bibfnamefont {S.}~\bibnamefont {Raghu}},\ }\bibfield  {title} {\bibinfo {title} {The hubbard model},\ }\href {https://doi.org/10.1146/annurev-conmatphys-031620-102024} {\bibfield  {journal} {\bibinfo  {journal} {Annual Review of Condensed Matter Physics}\ }\textbf {\bibinfo {volume} {13}},\ \bibinfo {pages} {239} (\bibinfo {year} {2022})},\ \Eprint {https://arxiv.org/abs/2103.12097} {arXiv:2103.12097 [cond-mat.str-el]} \BibitemShut {NoStop}%
\bibitem [{\citenamefont {Dagotto}(1994)}]{Dagotto1994}%
  \BibitemOpen
  \bibfield  {author} {\bibinfo {author} {\bibfnamefont {E.}~\bibnamefont {Dagotto}},\ }\bibfield  {title} {\bibinfo {title} {Correlated electrons in high-temperature superconductors},\ }\href {https://doi.org/10.1103/RevModPhys.66.763} {\bibfield  {journal} {\bibinfo  {journal} {Reviews of Modern Physics}\ }\textbf {\bibinfo {volume} {66}},\ \bibinfo {pages} {763} (\bibinfo {year} {1994})},\ \Eprint {https://arxiv.org/abs/cond-mat/9311013} {arXiv:cond-mat/9311013} \BibitemShut {NoStop}%
\bibitem [{\citenamefont {Lee}\ \emph {et~al.}(2006)\citenamefont {Lee}, \citenamefont {Nagaosa},\ and\ \citenamefont {Wen}}]{LeeNagaosaWen2006}%
  \BibitemOpen
  \bibfield  {author} {\bibinfo {author} {\bibfnamefont {P.~A.}\ \bibnamefont {Lee}}, \bibinfo {author} {\bibfnamefont {N.}~\bibnamefont {Nagaosa}},\ and\ \bibinfo {author} {\bibfnamefont {X.-G.}\ \bibnamefont {Wen}},\ }\bibfield  {title} {\bibinfo {title} {Doping a {Mott} insulator: Physics of high-temperature superconductivity},\ }\href {https://doi.org/10.1103/RevModPhys.78.17} {\bibfield  {journal} {\bibinfo  {journal} {Reviews of Modern Physics}\ }\textbf {\bibinfo {volume} {78}},\ \bibinfo {pages} {17} (\bibinfo {year} {2006})}\BibitemShut {NoStop}%
\bibitem [{\citenamefont {Qin}\ \emph {et~al.}(2022)\citenamefont {Qin}, \citenamefont {Sch{"a}fer}, \citenamefont {Andergassen}, \citenamefont {Corboz},\ and\ \citenamefont {Gull}}]{QinSchaferAndergassenCorbozGull2022}%
  \BibitemOpen
  \bibfield  {author} {\bibinfo {author} {\bibfnamefont {M.}~\bibnamefont {Qin}}, \bibinfo {author} {\bibfnamefont {T.}~\bibnamefont {Sch{"a}fer}}, \bibinfo {author} {\bibfnamefont {S.}~\bibnamefont {Andergassen}}, \bibinfo {author} {\bibfnamefont {P.}~\bibnamefont {Corboz}},\ and\ \bibinfo {author} {\bibfnamefont {E.}~\bibnamefont {Gull}},\ }\bibfield  {title} {\bibinfo {title} {The hubbard model: A computational perspective},\ }\href {https://doi.org/10.1146/annurev-conmatphys-090921-033948} {\bibfield  {journal} {\bibinfo  {journal} {Annual Review of Condensed Matter Physics}\ }\textbf {\bibinfo {volume} {13}},\ \bibinfo {pages} {275} (\bibinfo {year} {2022})},\ \Eprint {https://arxiv.org/abs/2104.00064} {arXiv:2104.00064 [cond-mat.str-el]} \BibitemShut {NoStop}%
\bibitem [{\citenamefont {Jakli{\v c}}\ and\ \citenamefont {Prelov{\v s}ek}(1994)}]{JaklicPrelovsek1994Conductivity}%
  \BibitemOpen
  \bibfield  {author} {\bibinfo {author} {\bibfnamefont {J.}~\bibnamefont {Jakli{\v c}}}\ and\ \bibinfo {author} {\bibfnamefont {P.}~\bibnamefont {Prelov{\v s}ek}},\ }\bibfield  {title} {\bibinfo {title} {Finite-temperature conductivity in the planar {$t$-$J$} model},\ }\href {https://doi.org/10.1103/PhysRevB.50.7129} {\bibfield  {journal} {\bibinfo  {journal} {Physical Review B}\ }\textbf {\bibinfo {volume} {50}},\ \bibinfo {pages} {7129} (\bibinfo {year} {1994})}\BibitemShut {NoStop}%
\bibitem [{\citenamefont {Jakli{\v c}}\ and\ \citenamefont {Prelov{\v s}ek}(1995)}]{JaklicPrelovsek1995ChargeDynamics}%
  \BibitemOpen
  \bibfield  {author} {\bibinfo {author} {\bibfnamefont {J.}~\bibnamefont {Jakli{\v c}}}\ and\ \bibinfo {author} {\bibfnamefont {P.}~\bibnamefont {Prelov{\v s}ek}},\ }\bibfield  {title} {\bibinfo {title} {Charge dynamics in the planar {$t$-$J$} model},\ }\href {https://doi.org/10.1103/PhysRevB.52.6903} {\bibfield  {journal} {\bibinfo  {journal} {Physical Review B}\ }\textbf {\bibinfo {volume} {52}},\ \bibinfo {pages} {6903} (\bibinfo {year} {1995})}\BibitemShut {NoStop}%
\bibitem [{\citenamefont {Zemlji{\v c}}\ and\ \citenamefont {Prelov{\v s}ek}(2005)}]{ZemljicPrelovsek2005Transport}%
  \BibitemOpen
  \bibfield  {author} {\bibinfo {author} {\bibfnamefont {M.~M.}\ \bibnamefont {Zemlji{\v c}}}\ and\ \bibinfo {author} {\bibfnamefont {P.}~\bibnamefont {Prelov{\v s}ek}},\ }\bibfield  {title} {\bibinfo {title} {Resistivity and optical conductivity of cuprates within the {$t$-$J$} model},\ }\href {https://doi.org/10.1103/PhysRevB.72.075108} {\bibfield  {journal} {\bibinfo  {journal} {Physical Review B}\ }\textbf {\bibinfo {volume} {72}},\ \bibinfo {pages} {075108} (\bibinfo {year} {2005})}\BibitemShut {NoStop}%
\bibitem [{\citenamefont {Huang}\ \emph {et~al.}(2019)\citenamefont {Huang}, \citenamefont {Sheppard}, \citenamefont {Moritz},\ and\ \citenamefont {Devereaux}}]{Huang2019}%
  \BibitemOpen
  \bibfield  {author} {\bibinfo {author} {\bibfnamefont {E.~W.}\ \bibnamefont {Huang}}, \bibinfo {author} {\bibfnamefont {R.}~\bibnamefont {Sheppard}}, \bibinfo {author} {\bibfnamefont {B.}~\bibnamefont {Moritz}},\ and\ \bibinfo {author} {\bibfnamefont {T.~P.}\ \bibnamefont {Devereaux}},\ }\bibfield  {title} {\bibinfo {title} {Strange metallicity in the doped {Hubbard} model},\ }\href {https://doi.org/10.1126/science.aau7063} {\bibfield  {journal} {\bibinfo  {journal} {Science}\ }\textbf {\bibinfo {volume} {366}},\ \bibinfo {pages} {987} (\bibinfo {year} {2019})}\BibitemShut {NoStop}%
\bibitem [{\citenamefont {Zhao}\ \emph {et~al.}(2025)\citenamefont {Zhao}, \citenamefont {Zhang}, \citenamefont {Wang}, \citenamefont {Ding}, \citenamefont {Liu}, \citenamefont {Moritz}, \citenamefont {Huang},\ and\ \citenamefont {Devereaux}}]{Zhao2025Emery}%
  \BibitemOpen
  \bibfield  {author} {\bibinfo {author} {\bibfnamefont {S.}~\bibnamefont {Zhao}}, \bibinfo {author} {\bibfnamefont {R.}~\bibnamefont {Zhang}}, \bibinfo {author} {\bibfnamefont {W.~O.}\ \bibnamefont {Wang}}, \bibinfo {author} {\bibfnamefont {J.~K.}\ \bibnamefont {Ding}}, \bibinfo {author} {\bibfnamefont {T.}~\bibnamefont {Liu}}, \bibinfo {author} {\bibfnamefont {B.}~\bibnamefont {Moritz}}, \bibinfo {author} {\bibfnamefont {E.~W.}\ \bibnamefont {Huang}},\ and\ \bibinfo {author} {\bibfnamefont {T.~P.}\ \bibnamefont {Devereaux}},\ }\bibfield  {title} {\bibinfo {title} {Enhanced superconducting correlations in the {Emery} model and its connections to strange metallic transport and normal state coherence},\ }\href {https://doi.org/10.1103/bbyx-gfwl} {\bibfield  {journal} {\bibinfo  {journal} {Physical Review B}\ }\textbf {\bibinfo {volume} {112}},\ \bibinfo {pages} {224513} (\bibinfo {year} {2025})}\BibitemShut {NoStop}%
\bibitem [{\citenamefont {Deng}\ \emph {et~al.}(2013)\citenamefont {Deng}, \citenamefont {Mravlje}, \citenamefont {Žitko}, \citenamefont {Ferrero}, \citenamefont {Kotliar},\ and\ \citenamefont {Georges}}]{Deng2013}%
  \BibitemOpen
  \bibfield  {author} {\bibinfo {author} {\bibfnamefont {X.}~\bibnamefont {Deng}}, \bibinfo {author} {\bibfnamefont {J.}~\bibnamefont {Mravlje}}, \bibinfo {author} {\bibfnamefont {R.}~\bibnamefont {Žitko}}, \bibinfo {author} {\bibfnamefont {M.}~\bibnamefont {Ferrero}}, \bibinfo {author} {\bibfnamefont {G.}~\bibnamefont {Kotliar}},\ and\ \bibinfo {author} {\bibfnamefont {A.}~\bibnamefont {Georges}},\ }\bibfield  {title} {\bibinfo {title} {How bad metals turn good: Spectroscopic signatures of resilient quasiparticles},\ }\href {https://doi.org/10.1103/PhysRevLett.110.086401} {\bibfield  {journal} {\bibinfo  {journal} {Physical Review Letters}\ }\textbf {\bibinfo {volume} {110}},\ \bibinfo {pages} {086401} (\bibinfo {year} {2013})}\BibitemShut {NoStop}%
\bibitem [{\citenamefont {Merino}\ and\ \citenamefont {McKenzie}(2000)}]{MerinoMcKenzie2000}%
  \BibitemOpen
  \bibfield  {author} {\bibinfo {author} {\bibfnamefont {J.}~\bibnamefont {Merino}}\ and\ \bibinfo {author} {\bibfnamefont {R.~H.}\ \bibnamefont {McKenzie}},\ }\bibfield  {title} {\bibinfo {title} {Transport properties of strongly correlated metals: A dynamical mean-field approach},\ }\href {https://doi.org/10.1103/PhysRevB.61.7996} {\bibfield  {journal} {\bibinfo  {journal} {Physical Review B}\ }\textbf {\bibinfo {volume} {61}},\ \bibinfo {pages} {7996} (\bibinfo {year} {2000})}\BibitemShut {NoStop}%
\bibitem [{\citenamefont {Vu{\v c}i{\v c}evi{\'c}}\ \emph {et~al.}(2015)\citenamefont {Vu{\v c}i{\v c}evi{\'c}}, \citenamefont {Tanaskovi{\'c}}, \citenamefont {Rozenberg},\ and\ \citenamefont {Dobrosavljevi{\'c}}}]{Vucicevic2015}%
  \BibitemOpen
  \bibfield  {author} {\bibinfo {author} {\bibfnamefont {J.}~\bibnamefont {Vu{\v c}i{\v c}evi{\'c}}}, \bibinfo {author} {\bibfnamefont {D.}~\bibnamefont {Tanaskovi{\'c}}}, \bibinfo {author} {\bibfnamefont {M.~J.}\ \bibnamefont {Rozenberg}},\ and\ \bibinfo {author} {\bibfnamefont {V.}~\bibnamefont {Dobrosavljevi{\'c}}},\ }\bibfield  {title} {\bibinfo {title} {Bad-metal behavior reveals {Mott} quantum criticality in doped {Hubbard} models},\ }\href {https://doi.org/10.1103/PhysRevLett.114.246402} {\bibfield  {journal} {\bibinfo  {journal} {Physical Review Letters}\ }\textbf {\bibinfo {volume} {114}},\ \bibinfo {pages} {246402} (\bibinfo {year} {2015})}\BibitemShut {NoStop}%
\bibitem [{\citenamefont {Vrani{\'c}}\ \emph {et~al.}(2020)\citenamefont {Vrani{\'c}}, \citenamefont {Vu{\v c}i{\v c}evi{\'c}}, \citenamefont {Kokalj},\ and\ \citenamefont {Žitko}}]{Vranic2020}%
  \BibitemOpen
  \bibfield  {author} {\bibinfo {author} {\bibfnamefont {A.}~\bibnamefont {Vrani{\'c}}}, \bibinfo {author} {\bibfnamefont {J.}~\bibnamefont {Vu{\v c}i{\v c}evi{\'c}}}, \bibinfo {author} {\bibfnamefont {J.}~\bibnamefont {Kokalj}},\ and\ \bibinfo {author} {\bibfnamefont {R.}~\bibnamefont {Žitko}},\ }\bibfield  {title} {\bibinfo {title} {Charge transport in the {Hubbard} model at high temperatures},\ }\href {https://doi.org/10.1103/PhysRevB.102.115142} {\bibfield  {journal} {\bibinfo  {journal} {Physical Review B}\ }\textbf {\bibinfo {volume} {102}},\ \bibinfo {pages} {115142} (\bibinfo {year} {2020})}\BibitemShut {NoStop}%
\bibitem [{\citenamefont {Jakli\v{c}}\ and\ \citenamefont {Prelov\v{s}ek}(2000)}]{JaklicPrelovsek2000}%
  \BibitemOpen
  \bibfield  {author} {\bibinfo {author} {\bibfnamefont {J.}~\bibnamefont {Jakli\v{c}}}\ and\ \bibinfo {author} {\bibfnamefont {P.}~\bibnamefont {Prelov\v{s}ek}},\ }\bibfield  {title} {\bibinfo {title} {Finite-temperature properties of doped antiferromagnets},\ }\href {https://doi.org/10.1080/000187300243381} {\bibfield  {journal} {\bibinfo  {journal} {Advances in Physics}\ }\textbf {\bibinfo {volume} {49}},\ \bibinfo {pages} {1} (\bibinfo {year} {2000})}\BibitemShut {NoStop}%
\bibitem [{\citenamefont {Kokalj}(2017)}]{Kokalj2017}%
  \BibitemOpen
  \bibfield  {author} {\bibinfo {author} {\bibfnamefont {J.}~\bibnamefont {Kokalj}},\ }\bibfield  {title} {\bibinfo {title} {Bad-metallic behavior of doped mott insulators},\ }\href {https://doi.org/10.1103/PhysRevB.95.041110} {\bibfield  {journal} {\bibinfo  {journal} {Physical Review B}\ }\textbf {\bibinfo {volume} {95}},\ \bibinfo {pages} {041110} (\bibinfo {year} {2017})}\BibitemShut {NoStop}%
\bibitem [{\citenamefont {Lindner}\ and\ \citenamefont {Auerbach}(2010)}]{LindnerAuerbach2010}%
  \BibitemOpen
  \bibfield  {author} {\bibinfo {author} {\bibfnamefont {N.~H.}\ \bibnamefont {Lindner}}\ and\ \bibinfo {author} {\bibfnamefont {A.}~\bibnamefont {Auerbach}},\ }\bibfield  {title} {\bibinfo {title} {Conductivity of hard core bosons: A paradigm of a bad metal},\ }\href {https://doi.org/10.1103/PhysRevB.81.054512} {\bibfield  {journal} {\bibinfo  {journal} {Physical Review B}\ }\textbf {\bibinfo {volume} {81}},\ \bibinfo {pages} {054512} (\bibinfo {year} {2010})}\BibitemShut {NoStop}%
\bibitem [{\citenamefont {Liu}\ \emph {et~al.}(2026)\citenamefont {Liu}, \citenamefont {Ma}, \citenamefont {Changlani}, \citenamefont {Phillips},\ and\ \citenamefont {Bernevig}}]{Liu2026Transport}%
  \BibitemOpen
  \bibfield  {author} {\bibinfo {author} {\bibfnamefont {S.}~\bibnamefont {Liu}}, \bibinfo {author} {\bibfnamefont {Y.}~\bibnamefont {Ma}}, \bibinfo {author} {\bibfnamefont {H.~J.}\ \bibnamefont {Changlani}}, \bibinfo {author} {\bibfnamefont {P.~W.}\ \bibnamefont {Phillips}},\ and\ \bibinfo {author} {\bibfnamefont {B.~A.}\ \bibnamefont {Bernevig}},\ }\href {https://doi.org/10.48550/arXiv.2604.02426} {\bibinfo {title} {Transport and temperature 1: Exact spectrum and resistivity for the one-dimensional infinite-{$U$} {Hubbard} model}} (\bibinfo {year} {2026}),\ \Eprint {https://arxiv.org/abs/2604.02426} {arXiv:2604.02426 [cond-mat.str-el]} \BibitemShut {NoStop}%
\bibitem [{\citenamefont {Mukerjee}\ \emph {et~al.}(2006)\citenamefont {Mukerjee}, \citenamefont {Oganesyan},\ and\ \citenamefont {Huse}}]{MukerjeeOganesyanHuse2006}%
  \BibitemOpen
  \bibfield  {author} {\bibinfo {author} {\bibfnamefont {S.}~\bibnamefont {Mukerjee}}, \bibinfo {author} {\bibfnamefont {V.}~\bibnamefont {Oganesyan}},\ and\ \bibinfo {author} {\bibfnamefont {D.}~\bibnamefont {Huse}},\ }\bibfield  {title} {\bibinfo {title} {Statistical theory of transport by strongly interacting lattice fermions},\ }\href {https://doi.org/10.1103/PhysRevB.73.035113} {\bibfield  {journal} {\bibinfo  {journal} {Physical Review B}\ }\textbf {\bibinfo {volume} {73}},\ \bibinfo {pages} {035113} (\bibinfo {year} {2006})}\BibitemShut {NoStop}%
\bibitem [{\citenamefont {Perepelitsky}\ \emph {et~al.}(2016)\citenamefont {Perepelitsky}, \citenamefont {Galatas}, \citenamefont {Mravlje}, \citenamefont {\v{Z}itko}, \citenamefont {Khatami}, \citenamefont {Shastry},\ and\ \citenamefont {Georges}}]{Perepelitsky2016}%
  \BibitemOpen
  \bibfield  {author} {\bibinfo {author} {\bibfnamefont {E.}~\bibnamefont {Perepelitsky}}, \bibinfo {author} {\bibfnamefont {A.}~\bibnamefont {Galatas}}, \bibinfo {author} {\bibfnamefont {J.}~\bibnamefont {Mravlje}}, \bibinfo {author} {\bibfnamefont {R.}~\bibnamefont {\v{Z}itko}}, \bibinfo {author} {\bibfnamefont {E.}~\bibnamefont {Khatami}}, \bibinfo {author} {\bibfnamefont {B.~S.}\ \bibnamefont {Shastry}},\ and\ \bibinfo {author} {\bibfnamefont {A.}~\bibnamefont {Georges}},\ }\bibfield  {title} {\bibinfo {title} {Transport and optical conductivity in the hubbard model: A high-temperature expansion perspective},\ }\href {https://doi.org/10.1103/PhysRevB.94.235115} {\bibfield  {journal} {\bibinfo  {journal} {Physical Review B}\ }\textbf {\bibinfo {volume} {94}},\ \bibinfo {pages} {235115} (\bibinfo {year} {2016})}\BibitemShut {NoStop}%
\bibitem [{\citenamefont {Nagaoka}(1966)}]{Nagaoka1966}%
  \BibitemOpen
  \bibfield  {author} {\bibinfo {author} {\bibfnamefont {Y.}~\bibnamefont {Nagaoka}},\ }\bibfield  {title} {\bibinfo {title} {Ferromagnetism in a narrow, almost half-filled $s$ band},\ }\href {https://doi.org/10.1103/PhysRev.147.392} {\bibfield  {journal} {\bibinfo  {journal} {Physical Review}\ }\textbf {\bibinfo {volume} {147}},\ \bibinfo {pages} {392} (\bibinfo {year} {1966})}\BibitemShut {NoStop}%
\bibitem [{\citenamefont {Liu}\ \emph {et~al.}(2012)\citenamefont {Liu}, \citenamefont {Yao}, \citenamefont {Berg}, \citenamefont {White},\ and\ \citenamefont {Kivelson}}]{LiuYaoBergWhiteKivelson2012}%
  \BibitemOpen
  \bibfield  {author} {\bibinfo {author} {\bibfnamefont {L.}~\bibnamefont {Liu}}, \bibinfo {author} {\bibfnamefont {H.}~\bibnamefont {Yao}}, \bibinfo {author} {\bibfnamefont {E.}~\bibnamefont {Berg}}, \bibinfo {author} {\bibfnamefont {S.~R.}\ \bibnamefont {White}},\ and\ \bibinfo {author} {\bibfnamefont {S.~A.}\ \bibnamefont {Kivelson}},\ }\bibfield  {title} {\bibinfo {title} {Phases of the infinite {$U$} {Hubbard} model on square lattices},\ }\href {https://doi.org/10.1103/PhysRevLett.108.126406} {\bibfield  {journal} {\bibinfo  {journal} {Physical Review Letters}\ }\textbf {\bibinfo {volume} {108}},\ \bibinfo {pages} {126406} (\bibinfo {year} {2012})}\BibitemShut {NoStop}%
\bibitem [{\citenamefont {Andreev}(1978)}]{Andreev1978}%
  \BibitemOpen
  \bibfield  {author} {\bibinfo {author} {\bibfnamefont {A.~F.}\ \bibnamefont {Andreev}},\ }\bibfield  {title} {\bibinfo {title} {Thermodynamics of liquids below the {Debye} temperature},\ }\href@noop {} {\bibfield  {journal} {\bibinfo  {journal} {JETP Letters}\ }\textbf {\bibinfo {volume} {28}},\ \bibinfo {pages} {556} (\bibinfo {year} {1978})},\ \bibinfo {note} {pis'ma Zh. Eksp. Teor. Fiz. 28, 603 (1978)}\BibitemShut {NoStop}%
\bibitem [{\citenamefont {Andreev}\ and\ \citenamefont {Kosevich}(1979)}]{AndreevKosevich1979}%
  \BibitemOpen
  \bibfield  {author} {\bibinfo {author} {\bibfnamefont {A.~F.}\ \bibnamefont {Andreev}}\ and\ \bibinfo {author} {\bibfnamefont {Y.~A.}\ \bibnamefont {Kosevich}},\ }\bibfield  {title} {\bibinfo {title} {Kinetic phenomena in semiquantum liquids},\ }\href@noop {} {\bibfield  {journal} {\bibinfo  {journal} {Soviet Physics JETP}\ }\textbf {\bibinfo {volume} {50}},\ \bibinfo {pages} {1218} (\bibinfo {year} {1979})},\ \bibinfo {note} {zh. Eksp. Teor. Fiz. 77, 2518 (1979)}\BibitemShut {NoStop}%
\bibitem [{\citenamefont {Betts}\ \emph {et~al.}(1999)\citenamefont {Betts}, \citenamefont {Lin},\ and\ \citenamefont {Flynn}}]{Betts1999}%
  \BibitemOpen
  \bibfield  {author} {\bibinfo {author} {\bibfnamefont {D.~D.}\ \bibnamefont {Betts}}, \bibinfo {author} {\bibfnamefont {H.~Q.}\ \bibnamefont {Lin}},\ and\ \bibinfo {author} {\bibfnamefont {J.~S.}\ \bibnamefont {Flynn}},\ }\bibfield  {title} {\bibinfo {title} {Improved finite-lattice estimates of the properties of two quantum spin models on the infinite square lattice},\ }\href@noop {} {\bibfield  {journal} {\bibinfo  {journal} {Canadian Journal of Physics}\ }\textbf {\bibinfo {volume} {77}},\ \bibinfo {pages} {353} (\bibinfo {year} {1999})}\BibitemShut {NoStop}%
\bibitem [{\citenamefont {Riera}\ and\ \citenamefont {Young}(1989)}]{RieraYoung1989}%
  \BibitemOpen
  \bibfield  {author} {\bibinfo {author} {\bibfnamefont {J.~A.}\ \bibnamefont {Riera}}\ and\ \bibinfo {author} {\bibfnamefont {A.~P.}\ \bibnamefont {Young}},\ }\bibfield  {title} {\bibinfo {title} {Ferromagnetism in the one-band {Hubbard} model},\ }\href {https://doi.org/10.1103/PhysRevB.40.5285} {\bibfield  {journal} {\bibinfo  {journal} {Physical Review B}\ }\textbf {\bibinfo {volume} {40}},\ \bibinfo {pages} {5285} (\bibinfo {year} {1989})}\BibitemShut {NoStop}%
\bibitem [{\citenamefont {Brinkman}\ and\ \citenamefont {Rice}(1970)}]{BrinkmanRice1970}%
  \BibitemOpen
  \bibfield  {author} {\bibinfo {author} {\bibfnamefont {W.~F.}\ \bibnamefont {Brinkman}}\ and\ \bibinfo {author} {\bibfnamefont {T.~M.}\ \bibnamefont {Rice}},\ }\bibfield  {title} {\bibinfo {title} {Single-particle excitations in magnetic insulators},\ }\href {https://doi.org/10.1103/PhysRevB.2.1324} {\bibfield  {journal} {\bibinfo  {journal} {Physical Review B}\ }\textbf {\bibinfo {volume} {2}},\ \bibinfo {pages} {1324} (\bibinfo {year} {1970})}\BibitemShut {NoStop}%
\bibitem [{\citenamefont {Vu{\v c}i{\v c}evi{\'c}}\ \emph {et~al.}(2019)\citenamefont {Vu{\v c}i{\v c}evi{\'c}}, \citenamefont {Kokalj}, \citenamefont {{\v Z}itko}, \citenamefont {Wentzell}, \citenamefont {Tanaskovi{\'c}},\ and\ \citenamefont {Mravlje}}]{VucicevicKokaljZitko2019Vertex}%
  \BibitemOpen
  \bibfield  {author} {\bibinfo {author} {\bibfnamefont {J.}~\bibnamefont {Vu{\v c}i{\v c}evi{\'c}}}, \bibinfo {author} {\bibfnamefont {J.}~\bibnamefont {Kokalj}}, \bibinfo {author} {\bibfnamefont {R.}~\bibnamefont {{\v Z}itko}}, \bibinfo {author} {\bibfnamefont {N.}~\bibnamefont {Wentzell}}, \bibinfo {author} {\bibfnamefont {D.}~\bibnamefont {Tanaskovi{\'c}}},\ and\ \bibinfo {author} {\bibfnamefont {J.}~\bibnamefont {Mravlje}},\ }\bibfield  {title} {\bibinfo {title} {Conductivity in the square lattice hubbard model at high temperatures: Importance of vertex corrections},\ }\href {https://doi.org/10.1103/PhysRevLett.123.036601} {\bibfield  {journal} {\bibinfo  {journal} {Physical Review Letters}\ }\textbf {\bibinfo {volume} {123}},\ \bibinfo {pages} {036601} (\bibinfo {year} {2019})}\BibitemShut {NoStop}%
\bibitem [{\citenamefont {Koepsell}\ \emph {et~al.}(2019)\citenamefont {Koepsell}, \citenamefont {Vijayan}, \citenamefont {Sompet}, \citenamefont {Grusdt}, \citenamefont {Hilker}, \citenamefont {Demler}, \citenamefont {Salomon}, \citenamefont {Bloch},\ and\ \citenamefont {Gross}}]{Koepsell2019}%
  \BibitemOpen
  \bibfield  {author} {\bibinfo {author} {\bibfnamefont {J.}~\bibnamefont {Koepsell}}, \bibinfo {author} {\bibfnamefont {J.}~\bibnamefont {Vijayan}}, \bibinfo {author} {\bibfnamefont {P.}~\bibnamefont {Sompet}}, \bibinfo {author} {\bibfnamefont {F.}~\bibnamefont {Grusdt}}, \bibinfo {author} {\bibfnamefont {T.~A.}\ \bibnamefont {Hilker}}, \bibinfo {author} {\bibfnamefont {E.}~\bibnamefont {Demler}}, \bibinfo {author} {\bibfnamefont {G.}~\bibnamefont {Salomon}}, \bibinfo {author} {\bibfnamefont {I.}~\bibnamefont {Bloch}},\ and\ \bibinfo {author} {\bibfnamefont {C.}~\bibnamefont {Gross}},\ }\bibfield  {title} {\bibinfo {title} {Imaging magnetic polarons in the doped fermi--hubbard model},\ }\href {https://doi.org/10.1038/s41586-019-1463-1} {\bibfield  {journal} {\bibinfo  {journal} {Nature}\ }\textbf {\bibinfo {volume} {572}},\ \bibinfo {pages} {358} (\bibinfo {year} {2019})}\BibitemShut {NoStop}%
\bibitem [{\citenamefont {Ji}\ \emph {et~al.}(2021)\citenamefont {Ji}, \citenamefont {Xu}, \citenamefont {Kendrick}, \citenamefont {Chiu}, \citenamefont {Br{\"u}ggenj{\"u}rgen}, \citenamefont {Greif}, \citenamefont {Bohrdt}, \citenamefont {Grusdt}, \citenamefont {Demler}, \citenamefont {Lebrat},\ and\ \citenamefont {Greiner}}]{Ji2021MagneticPolaron}%
  \BibitemOpen
  \bibfield  {author} {\bibinfo {author} {\bibfnamefont {G.}~\bibnamefont {Ji}}, \bibinfo {author} {\bibfnamefont {M.}~\bibnamefont {Xu}}, \bibinfo {author} {\bibfnamefont {L.~H.}\ \bibnamefont {Kendrick}}, \bibinfo {author} {\bibfnamefont {C.~S.}\ \bibnamefont {Chiu}}, \bibinfo {author} {\bibfnamefont {J.~C.}\ \bibnamefont {Br{\"u}ggenj{\"u}rgen}}, \bibinfo {author} {\bibfnamefont {D.}~\bibnamefont {Greif}}, \bibinfo {author} {\bibfnamefont {A.}~\bibnamefont {Bohrdt}}, \bibinfo {author} {\bibfnamefont {F.}~\bibnamefont {Grusdt}}, \bibinfo {author} {\bibfnamefont {E.}~\bibnamefont {Demler}}, \bibinfo {author} {\bibfnamefont {M.}~\bibnamefont {Lebrat}},\ and\ \bibinfo {author} {\bibfnamefont {M.}~\bibnamefont {Greiner}},\ }\bibfield  {title} {\bibinfo {title} {Coupling a mobile hole to an antiferromagnetic spin background: Transient dynamics of a magnetic polaron},\ }\href {https://doi.org/10.1103/PhysRevX.11.021022} {\bibfield  {journal} {\bibinfo  {journal} {Phys. Rev. X}\ }\textbf {\bibinfo {volume} {11}},\
  \bibinfo {pages} {021022} (\bibinfo {year} {2021})}\BibitemShut {NoStop}%
\bibitem [{\citenamefont {Koepsell}\ \emph {et~al.}(2021)\citenamefont {Koepsell}, \citenamefont {Bourgund}, \citenamefont {Sompet}, \citenamefont {Hirthe}, \citenamefont {Bohrdt}, \citenamefont {Wang}, \citenamefont {Grusdt}, \citenamefont {Demler}, \citenamefont {Salomon}, \citenamefont {Gross},\ and\ \citenamefont {Bloch}}]{Koepsell2021}%
  \BibitemOpen
  \bibfield  {author} {\bibinfo {author} {\bibfnamefont {J.}~\bibnamefont {Koepsell}}, \bibinfo {author} {\bibfnamefont {D.}~\bibnamefont {Bourgund}}, \bibinfo {author} {\bibfnamefont {P.}~\bibnamefont {Sompet}}, \bibinfo {author} {\bibfnamefont {S.}~\bibnamefont {Hirthe}}, \bibinfo {author} {\bibfnamefont {A.}~\bibnamefont {Bohrdt}}, \bibinfo {author} {\bibfnamefont {Y.}~\bibnamefont {Wang}}, \bibinfo {author} {\bibfnamefont {F.}~\bibnamefont {Grusdt}}, \bibinfo {author} {\bibfnamefont {E.}~\bibnamefont {Demler}}, \bibinfo {author} {\bibfnamefont {G.}~\bibnamefont {Salomon}}, \bibinfo {author} {\bibfnamefont {C.}~\bibnamefont {Gross}},\ and\ \bibinfo {author} {\bibfnamefont {I.}~\bibnamefont {Bloch}},\ }\bibfield  {title} {\bibinfo {title} {Microscopic evolution of doped {Mott} insulators from polaronic metal to {Fermi} liquid},\ }\href {https://doi.org/10.1126/science.abe7165} {\bibfield  {journal} {\bibinfo  {journal} {Science}\ }\textbf {\bibinfo {volume} {374}},\ \bibinfo {pages} {82} (\bibinfo {year}
  {2021})}\BibitemShut {NoStop}%
\bibitem [{\citenamefont {Lebrat}\ \emph {et~al.}(2024)\citenamefont {Lebrat}, \citenamefont {Xu}, \citenamefont {Kendrick}, \citenamefont {Kale}, \citenamefont {Gang}, \citenamefont {Seetharaman}, \citenamefont {Morera}, \citenamefont {Khatami}, \citenamefont {Demler},\ and\ \citenamefont {Greiner}}]{Lebrat2024}%
  \BibitemOpen
  \bibfield  {author} {\bibinfo {author} {\bibfnamefont {M.}~\bibnamefont {Lebrat}}, \bibinfo {author} {\bibfnamefont {M.}~\bibnamefont {Xu}}, \bibinfo {author} {\bibfnamefont {L.~H.}\ \bibnamefont {Kendrick}}, \bibinfo {author} {\bibfnamefont {A.}~\bibnamefont {Kale}}, \bibinfo {author} {\bibfnamefont {Y.}~\bibnamefont {Gang}}, \bibinfo {author} {\bibfnamefont {P.}~\bibnamefont {Seetharaman}}, \bibinfo {author} {\bibfnamefont {I.}~\bibnamefont {Morera}}, \bibinfo {author} {\bibfnamefont {E.}~\bibnamefont {Khatami}}, \bibinfo {author} {\bibfnamefont {E.}~\bibnamefont {Demler}},\ and\ \bibinfo {author} {\bibfnamefont {M.}~\bibnamefont {Greiner}},\ }\bibfield  {title} {\bibinfo {title} {Observation of {Nagaoka} polarons in a {Fermi--Hubbard} quantum simulator},\ }\href {https://doi.org/10.1038/s41586-024-07272-9} {\bibfield  {journal} {\bibinfo  {journal} {Nature}\ }\textbf {\bibinfo {volume} {629}},\ \bibinfo {pages} {317} (\bibinfo {year} {2024})}\BibitemShut {NoStop}%
\bibitem [{\citenamefont {Prichard}\ \emph {et~al.}(2024)\citenamefont {Prichard}, \citenamefont {Spar}, \citenamefont {Morera}, \citenamefont {Demler}, \citenamefont {Yan},\ and\ \citenamefont {Bakr}}]{Prichard2024}%
  \BibitemOpen
  \bibfield  {author} {\bibinfo {author} {\bibfnamefont {M.~L.}\ \bibnamefont {Prichard}}, \bibinfo {author} {\bibfnamefont {B.~M.}\ \bibnamefont {Spar}}, \bibinfo {author} {\bibfnamefont {I.}~\bibnamefont {Morera}}, \bibinfo {author} {\bibfnamefont {E.}~\bibnamefont {Demler}}, \bibinfo {author} {\bibfnamefont {Z.~Z.}\ \bibnamefont {Yan}},\ and\ \bibinfo {author} {\bibfnamefont {W.~S.}\ \bibnamefont {Bakr}},\ }\bibfield  {title} {\bibinfo {title} {Directly imaging spin polarons in a kinetically frustrated {Hubbard} system},\ }\href {https://doi.org/10.1038/s41586-024-07356-6} {\bibfield  {journal} {\bibinfo  {journal} {Nature}\ }\textbf {\bibinfo {volume} {629}},\ \bibinfo {pages} {323} (\bibinfo {year} {2024})}\BibitemShut {NoStop}%
\bibitem [{\citenamefont {Xu}\ \emph {et~al.}(2019)\citenamefont {Xu}, \citenamefont {McGehee}, \citenamefont {Morong},\ and\ \citenamefont {DeMarco}}]{Xu2019}%
  \BibitemOpen
  \bibfield  {author} {\bibinfo {author} {\bibfnamefont {W.}~\bibnamefont {Xu}}, \bibinfo {author} {\bibfnamefont {W.~R.}\ \bibnamefont {McGehee}}, \bibinfo {author} {\bibfnamefont {W.~N.}\ \bibnamefont {Morong}},\ and\ \bibinfo {author} {\bibfnamefont {B.}~\bibnamefont {DeMarco}},\ }\bibfield  {title} {\bibinfo {title} {Bad-metal relaxation dynamics in a {Fermi} lattice gas},\ }\href {https://doi.org/10.1038/s41467-019-09526-x} {\bibfield  {journal} {\bibinfo  {journal} {Nature Communications}\ }\textbf {\bibinfo {volume} {10}},\ \bibinfo {pages} {1588} (\bibinfo {year} {2019})}\BibitemShut {NoStop}%
\bibitem [{\citenamefont {Spivak}\ \emph {et~al.}(2010)\citenamefont {Spivak}, \citenamefont {Kravchenko}, \citenamefont {Kivelson},\ and\ \citenamefont {Gao}}]{Spivak2010}%
  \BibitemOpen
  \bibfield  {author} {\bibinfo {author} {\bibfnamefont {B.}~\bibnamefont {Spivak}}, \bibinfo {author} {\bibfnamefont {S.~V.}\ \bibnamefont {Kravchenko}}, \bibinfo {author} {\bibfnamefont {S.~A.}\ \bibnamefont {Kivelson}},\ and\ \bibinfo {author} {\bibfnamefont {X.~P.~A.}\ \bibnamefont {Gao}},\ }\bibfield  {title} {\bibinfo {title} {Colloquium: Transport in strongly correlated two dimensional electron fluids},\ }\href {https://doi.org/10.1103/RevModPhys.82.1743} {\bibfield  {journal} {\bibinfo  {journal} {Rev. Mod. Phys.}\ }\textbf {\bibinfo {volume} {82}},\ \bibinfo {pages} {1743} (\bibinfo {year} {2010})}\BibitemShut {NoStop}%
\bibitem [{\citenamefont {Hebard}\ \emph {et~al.}(1993)\citenamefont {Hebard}, \citenamefont {Palstra}, \citenamefont {Haddon},\ and\ \citenamefont {Fleming}}]{Hebard1993}%
  \BibitemOpen
  \bibfield  {author} {\bibinfo {author} {\bibfnamefont {A.~F.}\ \bibnamefont {Hebard}}, \bibinfo {author} {\bibfnamefont {T.~T.~M.}\ \bibnamefont {Palstra}}, \bibinfo {author} {\bibfnamefont {R.~C.}\ \bibnamefont {Haddon}},\ and\ \bibinfo {author} {\bibfnamefont {R.~M.}\ \bibnamefont {Fleming}},\ }\bibfield  {title} {\bibinfo {title} {Absence of saturation in the normal-state resistivity of thin films of {K$_3$C$_{60}$} and {Rb$_3$C$_{60}$}},\ }\href {https://doi.org/10.1103/PhysRevB.48.9945} {\bibfield  {journal} {\bibinfo  {journal} {Phys. Rev. B}\ }\textbf {\bibinfo {volume} {48}},\ \bibinfo {pages} {9945} (\bibinfo {year} {1993})}\BibitemShut {NoStop}%
\bibitem [{\citenamefont {Gunnarsson}(1997)}]{Gunnarsson1997}%
  \BibitemOpen
  \bibfield  {author} {\bibinfo {author} {\bibfnamefont {O.}~\bibnamefont {Gunnarsson}},\ }\bibfield  {title} {\bibinfo {title} {Superconductivity in fullerides},\ }\href {https://doi.org/10.1103/RevModPhys.69.575} {\bibfield  {journal} {\bibinfo  {journal} {Rev. Mod. Phys.}\ }\textbf {\bibinfo {volume} {69}},\ \bibinfo {pages} {575} (\bibinfo {year} {1997})}\BibitemShut {NoStop}%
\bibitem [{\citenamefont {{NVIDIA Corporation}}()}]{NVIDIA_cuSOLVERMp}%
  \BibitemOpen
  \bibfield  {author} {\bibinfo {author} {\bibnamefont {{NVIDIA Corporation}}},\ }\href {https://docs.nvidia.com/cuda/cusolvermp/} {\emph {\bibinfo {title} {{cuSOLVERMp}: A High-Performance CUDA Library for Distributed Dense Linear Algebra}}},\ \bibinfo {organization} {NVIDIA Corporation},\ \bibinfo {note} {version 0.8.0 (CUDA 12), accessed August 1, 2026}\BibitemShut {NoStop}%
\bibitem [{\citenamefont {Kohn}(1964)}]{Kohn1964}%
  \BibitemOpen
  \bibfield  {author} {\bibinfo {author} {\bibfnamefont {W.}~\bibnamefont {Kohn}},\ }\bibfield  {title} {\bibinfo {title} {Theory of the insulating state},\ }\href@noop {} {\bibfield  {journal} {\bibinfo  {journal} {Physical Review}\ }\textbf {\bibinfo {volume} {133}},\ \bibinfo {pages} {A171} (\bibinfo {year} {1964})}\BibitemShut {NoStop}%
\bibitem [{\citenamefont {Poilblanc}(1991)}]{Poilblanc1991}%
  \BibitemOpen
  \bibfield  {author} {\bibinfo {author} {\bibfnamefont {D.}~\bibnamefont {Poilblanc}},\ }\bibfield  {title} {\bibinfo {title} {Twisted boundary conditions in cluster calculations of the optical conductivity in two-dimensional lattice models},\ }\href {https://doi.org/10.1103/PhysRevB.44.9562} {\bibfield  {journal} {\bibinfo  {journal} {Physical Review B}\ }\textbf {\bibinfo {volume} {44}},\ \bibinfo {pages} {9562} (\bibinfo {year} {1991})}\BibitemShut {NoStop}%
\bibitem [{\citenamefont {Zotos}\ \emph {et~al.}(1997)\citenamefont {Zotos}, \citenamefont {Naef},\ and\ \citenamefont {Prelov{\v{s}}ek}}]{ZotosNaefPrelovsek1997}%
  \BibitemOpen
  \bibfield  {author} {\bibinfo {author} {\bibfnamefont {X.}~\bibnamefont {Zotos}}, \bibinfo {author} {\bibfnamefont {F.}~\bibnamefont {Naef}},\ and\ \bibinfo {author} {\bibfnamefont {P.}~\bibnamefont {Prelov{\v{s}}ek}},\ }\bibfield  {title} {\bibinfo {title} {Transport and conservation laws},\ }\href {https://doi.org/10.1103/PhysRevB.55.11029} {\bibfield  {journal} {\bibinfo  {journal} {Phys. Rev. B}\ }\textbf {\bibinfo {volume} {55}},\ \bibinfo {pages} {11029} (\bibinfo {year} {1997})}\BibitemShut {NoStop}%
\bibitem [{\citenamefont {Prange}\ and\ \citenamefont {Kadanoff}(1964)}]{prange_transport_1964}%
  \BibitemOpen
  \bibfield  {author} {\bibinfo {author} {\bibfnamefont {R.~E.}\ \bibnamefont {Prange}}\ and\ \bibinfo {author} {\bibfnamefont {L.~P.}\ \bibnamefont {Kadanoff}},\ }\bibfield  {title} {{\selectlanguage {english}\bibinfo {title} {Transport {Theory} for {Electron}-{Phonon} {Interactions} in {Metals}}},\ }\href {https://doi.org/10.1103/PhysRev.134.A566} {\bibfield  {journal} {\bibinfo  {journal} {Physical Review}\ }\textbf {\bibinfo {volume} {134}},\ \bibinfo {pages} {A566} (\bibinfo {year} {1964})}\BibitemShut {NoStop}%
\bibitem [{\citenamefont {Mott}(1972)}]{mott1972}%
  \BibitemOpen
  \bibfield  {author} {\bibinfo {author} {\bibfnamefont {N.~F.}\ \bibnamefont {Mott}},\ }\bibfield  {title} {\bibinfo {title} {Conduction in non-crystalline systems {IX}. the minimum metallic conductivity},\ }\href@noop {} {\bibfield  {journal} {\bibinfo  {journal} {Philosophical Magazine}\ }\textbf {\bibinfo {volume} {26}},\ \bibinfo {pages} {1015} (\bibinfo {year} {1972})}\BibitemShut {NoStop}%
\bibitem [{\citenamefont {Brown}\ \emph {et~al.}(2020)\citenamefont {Brown}, \citenamefont {Guardado-Sanchez}, \citenamefont {Spar}, \citenamefont {Huang}, \citenamefont {Devereaux},\ and\ \citenamefont {Bakr}}]{brown2020arpes}%
  \BibitemOpen
  \bibfield  {author} {\bibinfo {author} {\bibfnamefont {P.~T.}\ \bibnamefont {Brown}}, \bibinfo {author} {\bibfnamefont {E.}~\bibnamefont {Guardado-Sanchez}}, \bibinfo {author} {\bibfnamefont {B.~M.}\ \bibnamefont {Spar}}, \bibinfo {author} {\bibfnamefont {E.~W.}\ \bibnamefont {Huang}}, \bibinfo {author} {\bibfnamefont {T.~P.}\ \bibnamefont {Devereaux}},\ and\ \bibinfo {author} {\bibfnamefont {W.~S.}\ \bibnamefont {Bakr}},\ }\bibfield  {title} {\bibinfo {title} {Angle-resolved photoemission spectroscopy of a {Fermi--Hubbard} system},\ }\href {https://doi.org/10.1038/s41567-019-0696-0} {\bibfield  {journal} {\bibinfo  {journal} {Nature Physics}\ }\textbf {\bibinfo {volume} {16}},\ \bibinfo {pages} {26} (\bibinfo {year} {2020})}\BibitemShut {NoStop}%
\end{thebibliography}%


\appendix

\section{Model, Hilbert space, symmetries, and boundary conditions}
\label{sec:model}

Here we briefly provide several basic definitions of the model and discussed observables and outline the structure of the code.

\subsection{Model and Hilbert space}
We study the infinite-$U$ Hubbard model on finite clusters of the square lattice,
\begin{equation}
H \;=\; -t \sum_{\langle ij\rangle,\sigma}
\left( \tilde c^{\dagger}_{i\sigma}\tilde c_{j\sigma} + \mathrm{h.c.} \right),
\qquad
\tilde c_{i\sigma} = c_{i\sigma}\,(1-n_{i\bar\sigma}),
\label{eq:H}
\end{equation}
where the Gutzwiller-projected operators $\tilde c_{i\sigma}$ enforce the
no-double-occupancy constraint, and the hole doping is $x = N_h/N$ with $N_h$
the number of holes and $N$ the number of sites.  For fixed electron number
$N_e = N - N_h$, the dimension of the constrained Hilbert space is
$\binom{N}{N_h}\,2^{N_e}$, i.e., the number of hole placements times the
number of spin configurations of the remaining singly occupied sites.

The symmetries used in the calculation are:
(i)~$U(1)$ charge conservation, $[H,\hat N_e]=0$;
(ii)~$SU(2)$ spin symmetry, $[H,\mathbf S]=0$, of which we exploit $S_z$
conservation together with the spin-flip degeneracy between $\pm S_z$
sectors, halving the number of blocks to be diagonalized;
(iii)~lattice translations, whose quantum numbers depend on the boundary
conditions as specified below.
Twists and staggering generically reduce the $C_4$ point group of the
untwisted square clusters to $C_2$ or break it entirely; we therefore do not
use point-group symmetries anywhere.
The Hamiltonian is block diagonal in $(N_e, S_z, \mathbf k)$, and each
symmetry block is fully diagonalized with dense eigensolvers, retaining the
complete many-body spectrum and all current matrix elements.  The block
structure and the largest dense-block dimensions are collected in
Table~\ref{tab:clusters}.

\subsection{Cluster geometries}
\begin{figure}[t!]
\centering
\includegraphics{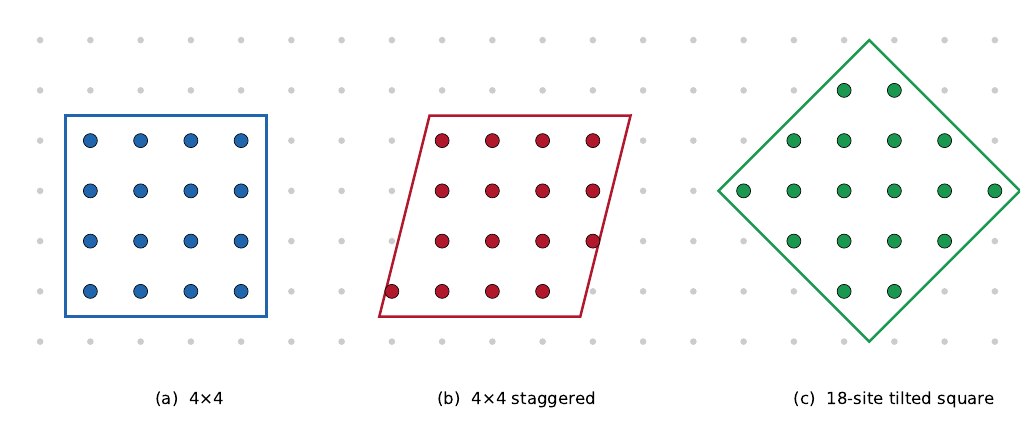}
\caption{\textbf{Cluster geometries.}
Representative finite clusters used in the exact-diagonalization
calculations: (\textbf{a})~the $4\times4$ square torus, spanned by
$\mathbf L_1=(4,0)$, $\mathbf L_2=(0,4)$; (\textbf{b})~the staggered
$4\times4$ torus ($s=1$), in which the $y$~boundary is reconnected with a
unit shift, $\mathbf L_2=(-s,4)$, modifying the allowed momenta according to
Eq.~(\ref{eq:kthetas}); (\textbf{c})~the 18-site tilted square Betts cluster,
$\mathbf L_1=(3,3)$, $\mathbf L_2=(3,-3)$. }
\label{fig:geometries}
\end{figure}

Three cluster geometries are used for the transport and thermodynamics
results (Fig.~\ref{fig:geometries}):
the $4\times4$ square torus, spanned by $\mathbf L_1=(4,0)$ and
$\mathbf L_2=(0,4)$;
the \emph{staggered} $4\times4$ torus ($s=1$), in which the $y$~boundary is
reconnected with a unit shift, $(x, N_y) \sim (x+s,\,0)$, corresponding to
$\mathbf L_2=(-s,4)$;
and the 18-site tilted (Betts) cluster~\cite{Betts1999}, spanned by
$\mathbf L_1=(3,3)$ and $\mathbf L_2=(3,-3)$.
The $4\times4$ torus is well known to be a pathological geometry at the
non-interacting level: its single-particle spectrum has a high degree of
fine-tuned degeneracy, which collapses many-body levels onto a few distinct
energies and enhances finite-size effects.  Tilting (in the spirit of the Betts construction) and
staggering modify the set of allowed momenta and lift these fine-tuned
degeneracies; comparing the $s=0$ and $s=1$ sectors of the $4\times4$
cluster, and the $4\times4$ against the 18-site cluster, therefore provides
independent finite-size diagnostics (App.~\ref{sec:fse}).
The variational caging analysis of App.~\ref{sec:caging} additionally uses a
$2\times8$ ladder with one hole.

\subsection{Twisted and staggered boundary conditions}
Twisted boundary conditions $\bm\theta = (\theta_x,\theta_y)$ are implemented
by attaching phases to the hopping matrix elements,
$t \to t\, e^{-i\bm\theta\cdot\bm\delta}$, with $\bm\delta$ the bond vector.
For the simple torus the allowed momenta are
\begin{equation}
\mathbf k(\bm\theta) =
\left( \frac{2\pi}{L_x} n_x + \theta_x,\;
       \frac{2\pi}{L_y} n_y + \theta_y \right),
\label{eq:ktheta}
\end{equation}
while on the staggered torus, $(x,N_y)\sim(x+s,0)$,
\begin{equation}
\mathbf k(\bm\theta, s) =
\left( \frac{2\pi}{L_x} n_x + \theta_x,\;
       \frac{2\pi}{L_y}\Big( n_y + \frac{n_x s}{L_x} \Big) + \theta_y \right).
\label{eq:kthetas}
\end{equation}
Equation~(\ref{eq:kthetas}) satisfies the boundary conditions of
the torus spanned by $\mathbf L_1 = (L_x,0)$, $\mathbf L_2=(-s,L_y)$:
$\mathcal T_x^{L_x} = 1$ and $\mathcal T_y^{L_y}\mathcal T_x^{-s} = 1$.
Because the twist enters only through the hopping phases, the eigenstates
are strictly periodic on the torus, and the translation eigenvalues involve
only the combination $\mathbf k - \bm\theta$.  The closure conditions
therefore require $(\mathbf k - \bm\theta)\cdot\mathbf L_{1,2} \in
2\pi\mathbb{Z}$, which Eq.~(\ref{eq:kthetas}) satisfies for every twist:
\begin{equation*}
(k_x-\theta_x)\,L_x = 2\pi n_x,
\qquad
(k_y-\theta_y)\,L_y - (k_x-\theta_x)\,s
 = 2\pi\Big( n_y + \frac{n_x s}{L_x} \Big) - \frac{2\pi n_x s}{L_x}
 = 2\pi n_y .
\end{equation*}
Staggering thus reshuffles the allowed momenta at fixed twist, so the
$s=0$ and $s=1$ sectors provide independent finite-size samples.

For twist averaging we use a $5\times5$ ``golden-ratio'' grid,
\begin{equation}
\theta_{x} \in \left\{ \frac{2\pi (a + \varphi)}{5} \right\}_{a=0}^{4},
\qquad
\theta_{y} \in \left\{ \frac{2\pi (b + \varphi)}{5} \right\}_{b=0}^{4},
\qquad \varphi = \frac{1+\sqrt5}{2},
\label{eq:goldengrid}
\end{equation}
i.e., 25 twists per boundary-condition sector.  The irrational offset
$\varphi$ keeps every twist away from high-symmetry commensurate points,
avoiding the fine-tuned degeneracies discussed above.  The same grid is used
for the $4\times4$ cluster in both boundary-condition sectors ($s=0$ and
$s=1$) and, in full, for the 18-site cluster (25 twists); the $x=3/16$
data shown in Fig.~\ref{fig:thermo} are for a single irrational twist.

\subsection{Numerical implementation}
\label{sec:implementation}

We use a custom Fortran exact-diagonalization code developed
for this work. The code diagonalizes
each dense symmetry block, and its structure is organized to make this
full diagonalization and the subsequent spectral sums feasible at the sizes
of Table~\ref{tab:clusters}.

Each $(N_e, S_z)$ sector is generated directly in the constrained
(no-double-occupancy) Fock basis, with the Hamiltonian and current operators
stored and applied sparsely.  When an operator maps one basis state to
another, the target index is found by binary search on a lexicographically
sorted key map, e.g. a standard sorted-basis lookup, which keeps operator assembly
near-linear in the sector dimension.  Translation symmetry, carrying the
attendant fermionic signs, then blocks each sector by lattice momentum
$\mathbf k(\bm\theta, s)$; together with particle number and $S_z$ (with the
$\pm S_z$ degeneracy) this leaves a set of independent dense blocks whose
largest dimension controls the cost and feasibility of the computation.

Every dense block is diagonalized in full on GPUs with the distributed multi-GPU
eigensolver NVIDIA \texttt{cuSOLVERMp}~\cite{NVIDIA_cuSOLVERMp}.  The eigenpairs
are then consumed in a single streaming pass, i.e., for each batch of eigenstates
the current and kinetic-energy matrix elements are formed by sparse scatters
and dense rotations, and their contributions to the
conductivity and to the thermodynamic traces are
accumulated on the fly, simultaneously for all temperatures, frequencies,
and broadenings.  The full spectral matrices are therefore never stored,
which is what keeps the memory footprint manageable at the largest block
sizes.

\begin{table}[t!]
\centering
\small
\begin{tabular}{lccccl}
\toprule
Cluster & $N_h$ & $x$ & $\dim \mathcal H_{N_e}$ & largest $(S_z,\mathbf k)$ block & twists / role \\
\midrule
$4\times4$ ($s=0$ and $s=1$) & 0 & 0    & $2^{16}$      & analytic, Eq.~(\ref{eq:chic}) & $\chic$ reference \\
$4\times4$ ($s=0$ and $s=1$) & 1 & 1/16 & 524{,}288     & 6{,}435   & 25 twists \\
$4\times4$ ($s=0$ and $s=1$) & 2 & 2/16 & 1{,}966{,}080 & 25{,}740  & 25 twists \\
$4\times4$                    & 3 & 3/16 & 4{,}587{,}520 & 60{,}060  & 1 irrational twist; thermodynamics only \\
18-site Betts                 & 0 & 0    & $2^{18}$      & analytic & $\chic$ reference \\
18-site Betts                 & 1 & 1/18 & 2{,}359{,}296 & 24{,}310  & 25 twists; untwisted for spin-bond correlator \\
18-site Betts                 & 2 & 2/18 & 10{,}027{,}008 & 109{,}430 & 1 irrational twist; thermodynamics only \\
$4\times5$                    & 1 & 1/20 & 10{,}485{,}760 & 92{,}378 & 25 twists \\
$2\times8$ ladder             & 1 & 1/16 & 524{,}288     & 6{,}435   & untwisted; caging analysis (App.~\ref{sec:caging}) \\
\bottomrule
\end{tabular}
\caption{\textbf{Clusters, dopings, and Hilbert-space dimensions.}
$\dim\mathcal H_{N_e} = \binom{N}{N_h} 2^{N_e}$ is the dimension of the
constrained Hilbert space at fixed hole number; the largest dense block is
the largest $(S_z,\mathbf k)$ sector actually diagonalized (spin-flip
degeneracy halves the number of $S_z$ blocks). Common numerical parameters:
temperature grid of 80 points, linear on $T\in[0.01,2]\,t$; frequency grid of
201 points up to $\omega = 10\,t$, densified at low frequency; broadenings
$\gamma/t \in \{0.005, 0.01, 0.02, 0.05, 0.1\}$.}
\label{tab:clusters}
\end{table}

\section{Thermodynamic observables}
\label{sec:thermo}

All thermodynamic quantities are computed from the full spectra of the
particle-number sectors.  For each sector $N_e$ and twist $\bm\theta$ the
canonical partition function and free energy are
\begin{equation}
Z_{N_e}(T,\bm\theta) = \sum_{n} e^{-E_n(\bm\theta)/T},
\qquad
F_{N_e}(T,\bm\theta) = -T \ln Z_{N_e}(T,\bm\theta),
\end{equation}
with the sums running over all eigenstates of the sector.  Unless stated
otherwise, sector free energies and observables are averaged over the twist
grid, Eq.~(\ref{eq:goldengrid}).  The entropy and specific heat per site are
obtained by numerical differentiation of the exact $F(T)$ and
$E(T) = \langle H\rangle_T$ on the temperature grid,
$S = -\partial F/\partial T /N$ and $C_V = \partial E/\partial T/N$.
The uniform spin susceptibility is computed as in the main text,
$\chis = (\langle S_z^2\rangle - \langle S_z\rangle^2)/(NT)$.

The charge compressibility is obtained from the discrete second difference
of the sector free energies with respect to the electron number,
\begin{equation}
\chic^{-1}(T)\;=\; N \Big[ F_{N_e+1}(T) - 2 F_{N_e}(T) + F_{N_e-1}(T) \Big],
\label{eq:chic}
\end{equation}
evaluated at the electron number $N_e$ corresponding to the doping of
interest.  For the single-hole dopings the
half-filled sector enters as one of the neighboring sectors; at $U=\infty$
it has no charge dynamics, so its free energy is exact,
$F_{N} = -NT\ln 2$ (free spins).  Concretely, $\chic(x{=}1/18)$ uses the
18-site sectors $N_e = 18, 17, 16$, $\chic(x{=}1/16)$ uses the $4\times4$
sectors $N_e = 16, 15, 14$, and $\chic(x{=}2/16)$ uses $N_e = 15, 14, 13$.
Each sector free energy is averaged over its available twists
(Table~\ref{tab:clusters}): all $4\times4$ sectors with $N_h \le 2$ and the
18-site one-hole sector use the full 25-twist grid, while the outermost
sectors of the $x=1/18$ and $x=2/16$ differences ($N_e = 16$ on the 18-site
cluster and $N_e = 13$ on the $4\times4$ cluster) are available for a single
twist.

\section{Optical conductivity and the dc limit}
\label{sec:optical}

\subsection{Kubo formula and broadening}
The longitudinal optical conductivity is computed from the Lehmann
representation,
\begin{equation}
\sigma_\mu(\omega)\;=\;\pi\,
\frac{1-e^{-\omega/T}}{\omega}
\sum_{n,m} \frac{e^{-E_n/T}}{Z}\,
\left|\langle m | J_\mu | n\rangle\right|^2
\delta\big(\omega - (E_m - E_n)\big),
\qquad \mu = x,y,
\label{eq:kubo}
\end{equation}
with $J_\mu$ the (projected) current operator, using every eigenstate and
every current matrix element of each symmetry block.  Here and throughout
the appendices, $\sigma(\omega)$ denotes the real (dissipative) part of the
conductivity, and the reported conductivity is the average of the two
diagonal components, $\sigma = (\sigma_x + \sigma_y)/2$ (the anisotropy is at most a few percent in the staggered-BC case).

\begin{figure}[t!]
\centering
\includegraphics[width=0.9\textwidth]{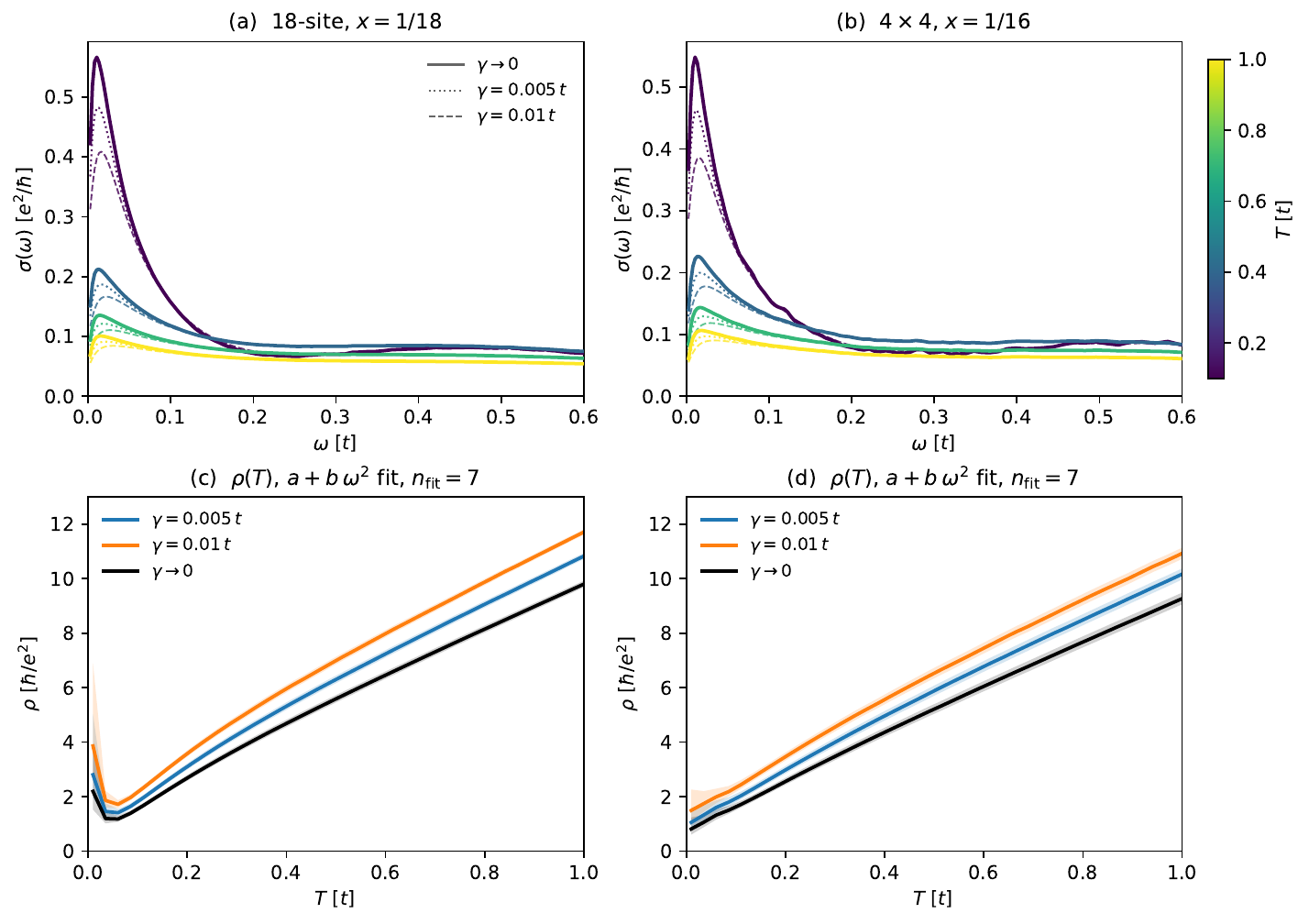}
\caption{\textbf{$\gamma\to0$ extrapolation of the optical conductivity and
 the dc resistivity.}  (\textbf{a}) and (\textbf{b}) Optical conductivity at representative
temperatures for the sampled broadenings $\gamma/t = 0.005, 0.01$, together with the $\gamma\to0$ extrapolation from the two
smallest values, for the 18-sites and $4\times4$ clusters, respectively.
(\textbf{c}) and (\textbf{d}) dc resistivity obtained by applying the extraction of
App.~\ref{sec:dcextraction} to the fixed-broadening spectra with
$\gamma = 0.005\,t$ and $0.01\,t$ and to the $\gamma\to0$-extrapolated
spectra. Twist error becomes significant below $T\lesssim 0.08t$, indicating increased sensitivity to finite size effects, similarly to the non-staggered/staggered comparison in the main text.}
\label{fig:gammaextrap}
\end{figure}

The delta functions
are broadened into Lorentzians,
$\delta(\omega-\Delta E) \to
\frac{1}{\pi}\,\gamma/[\gamma^2 + (\omega-\Delta E)^2]$,
and the calculation is repeated for
$\gamma/t \in \{0.005, 0.01, 0.02, 0.05, 0.1\}$.
The physically meaningful broadening window is bounded below by the
finite-size level spacing and above by the temperature,
$\Delta_{\mathrm{fs}} \lesssim \gamma \ll T$: for smaller $\gamma$ the
spectrum resolves into discrete peaks, while larger $\gamma$ artificially
redistributes spectral weight. In practice, to minimize the systematic loss of spectral weight caused by introducing broadening, we take the
$\gamma\to0$ limit by linear extrapolation in $\gamma$ from the two smallest
values, $\sigma(\omega,\gamma) = \sigma_0(\omega) + a_\omega \gamma + \mathcal{O}(\gamma^2)$. The results obtained are robust to other extrapolation schemes, e.g., using more broadening values or adding higher order terms in $\gamma$. Representative extrapolations for both clusters are shown in
Fig.~\ref{fig:gammaextrap}(a,b), and Fig.~\ref{fig:gammaextrap}(c,d) shows the
resulting dc resistivity obtained from the fixed-broadening spectra
alongside the $\gamma\to0$ extrapolation. In all cases we find that the $\gamma$ dependence is regular and enables a systematic extrapolation.

We also note that the residual resistivity obtained from extrapolating the $T$-linear resistivity in the semi-quantum regime to $T=0$ is reduced for larger systems (see Fig.~\ref{fig:rho_kappa_D}), which suggests it is a finite-size artifact.

\begin{figure}[H]
\centering
\includegraphics[width=\linewidth]{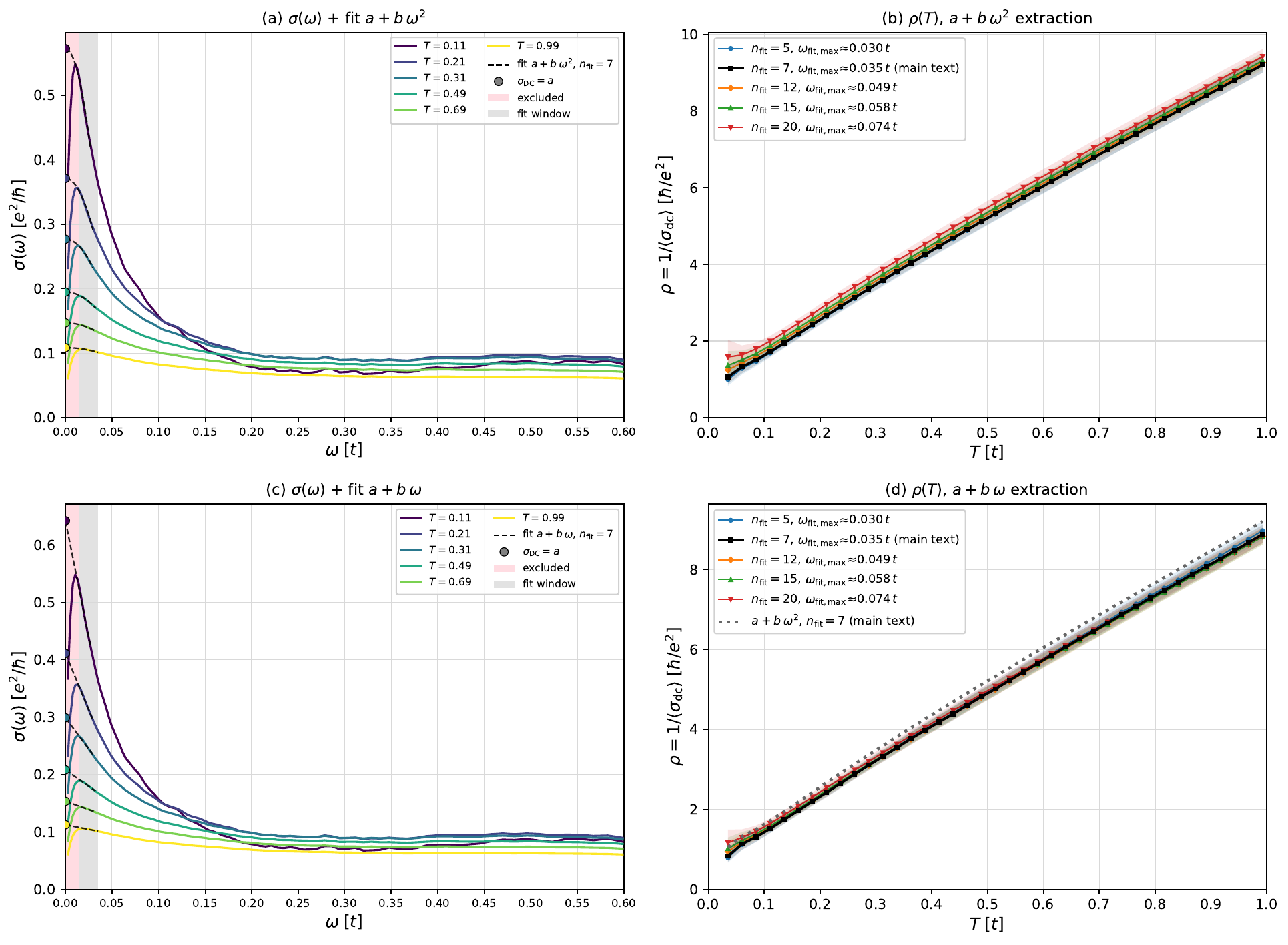}
\caption{\textbf{Extraction of the dc conductivity and its robustness
in representative case $4\times4$, $x=1/16$.}
(\textbf{a})~Twist-averaged, $\gamma\to0$-extrapolated $\sigma(\omega)$ at
representative temperatures.  The pink band marks the level-repulsion
downturn at $\omega<\omega_{\mathrm{peak}}$, excluded from the fit; the gray
band marks the fit window $[\omega_{\mathrm{peak}},\omega_{\mathrm{fit,max}}]$
used for the quadratic extrapolation $\sigma(\omega)=\sdc+b\,\omega^2$ (dashed
lines; circles mark the extracted $\sdc=a$ at $\omega=0$).
(\textbf{b})~Twist-averaged resistivity
$\rho(T)=1/\langle\sdc\rangle_{\bm\theta}$ from the quadratic fit for several
fit windows, labeled by their upper frequency
$\omega_{\mathrm{fit,max}}\approx0.03$--$0.07\,t$; the main-text choice is
$\omega_{\mathrm{fit,max}}\approx0.05\,t$.  Bands show
$1/(\langle\sdc\rangle\mp\mathrm{SEM})$ over the twists.
(\textbf{c},\textbf{d})~The same, using a \emph{linear} extrapolation
$\sigma(\omega)=\sdc+b\,\omega$; panel~(\textbf{d}) overlays the quadratic
main-text result (dotted).  The two fit forms differ by a modest offset
($\sim\!10$, largest near $0.08t$, below which twist variance grows rapidly and finite-size effects dominate) but give the same $T$-linear slope,
and within each form the fit window changes $\rho(T)$ by no more than about
the twist standard error for $T>\Tfs$.}
\label{fig:dcextraction}
\end{figure}

\subsection{Extraction of the dc limit}
\label{sec:dcextraction}
At the lowest frequencies the finite cluster deviates from the expected regular
$\omega\to0$ behavior of the thermodynamic limit as finite cluster artifacts depletes
the spectral weight in a narrow window near $\omega=0$, producing a downturn
of $\sigma(\omega)$ below a temperature-dependent scale
$\omega_{\mathrm{peak}}$~\cite{MukerjeeOganesyanHuse2006}.  We therefore extract $\sdc$ by
excluding the downturn window and fitting
\begin{equation}
\sigma(\omega)\;=\;\sdc+b\,\omega^2,
\qquad \omega\in[\omega_{\mathrm{peak}},\,\omega_{\mathrm{fit,max}}],
\label{eq:dcfit}
\end{equation}
where $\omega_{\mathrm{peak}}$ is the local maximum of the per-twist,
$\gamma\to0$-extrapolated $\sigma(\omega)$ near $\omega=0$, and the fit window
extends to $\omega_{\mathrm{fit,max}}\approx0.05\,t$ in the main text.  The extraction is performed
independently for each twist, and the resistivity is obtained from the
twist-averaged conductivity,
$\rho(T)=1/\langle\sdc(\bm\theta)\rangle_{\bm\theta}$, with the uncertainty
band propagated from the standard error of $\langle\sdc\rangle$, i.e.\
$\rho_{\pm}=1/(\langle\sdc\rangle\mp\mathrm{SEM})$.  The extraction can be done when the excluded window is small compared with the physical scales,
$\max(\omega_{\mathrm{peak}},\gamma)\ll\min(T,t)$, which is satisfied
throughout the reported temperature range.

Figure~\ref{fig:dcextraction} tests the stability of this procedure.  Varying
the fit window over $\omega_{\mathrm{fit,max}}\approx0.03$--$0.07\,t$
[Fig.~\ref{fig:dcextraction}(b)] changes $\rho(T)$ by no more than about the
twist standard error for $T>\Tfs$.  Replacing the quadratic fit by a linear
one, $\sigma(\omega)=\sdc+b\,\omega$ [Fig.~\ref{fig:dcextraction}(c,d)],
shifts $\rho(T)$ downward by a small constant amount but
leaves the $T$-linear slope essentially unchanged.  The characteristics of
the semi-quantum regime discussed in the main text, i.e., the magnitude of $\rho$,
its approximate $T$-linearity, and its slope, are thus insensitive both to
the fit window and to the fit form.

\subsection{More details on finite size effects}
\label{sec:fse}

\begin{figure}[H]
\centering
\includegraphics[width=\linewidth]{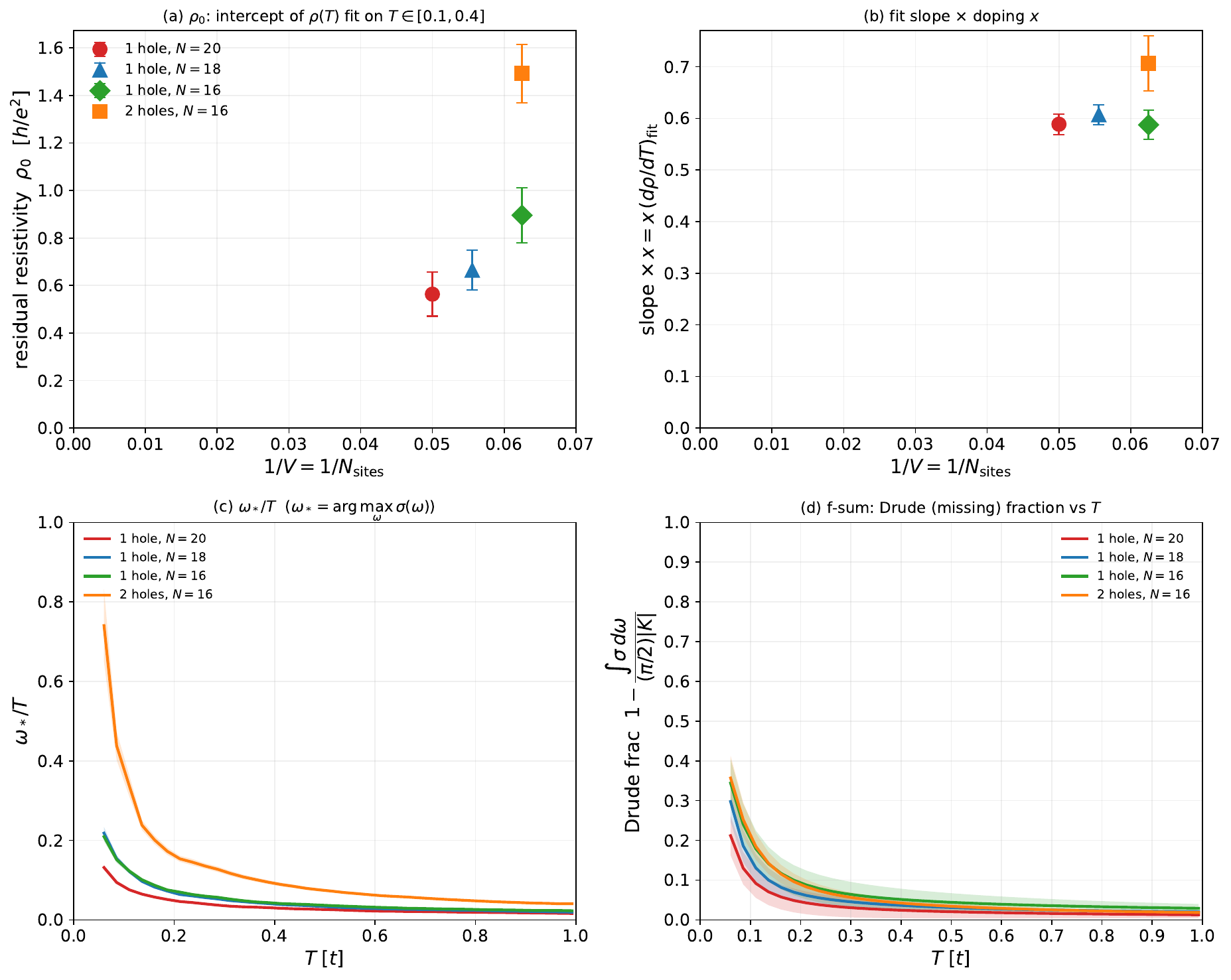}
\caption{\textbf{Finite-size diagnostics of charge transport.}
(A) Residual resistivity $\rho_0$ obtained from fits
$\rho(T)=\rho_0+AT$ over $0.1\leq T/t\leq0.4$, shown versus inverse system
size. (B) Corresponding slope $A$ multiplied by the doping $x$, testing the
expected low-doping scaling $\rho\sim T/x$. (C) Position $\omega^\ast$ of
the low-frequency maximum of $\sigma(\omega)$, normalized by $T$.
(D) Drude-weight fraction, defined as the spectral weight missing from the
regular optical conductivity. Results are shown for one hole on
$N=16,18,20$ clusters and two holes on $N=16$.}
\label{fig:residualetc}
\end{figure}

We further assess the finite-size dependence of the transport results in
Fig.~\ref{fig:residualetc}. Panel (a) shows the residual resistivity
$\rho_0$, obtained from a linear fit $\rho(T)=\rho_0+AT$ over the
temperature window indicated in the caption; the results are insensitive
to moderate changes of this window. In the one-hole sector, $\rho_0$
decreases systematically upon increasing the system size from $N=16$ to
$18$ and $20$. Although the doping simultaneously decreases along this
sequence, the systematic suppression with system size suggests that the
nonzero intercept is predominantly a finite-size effect, consistent with
its disappearance toward the thermodynamic limit. The two-hole $N=16$
result has a substantially larger intercept, indicating stronger
finite-size corrections in this sector.

In contrast, the slope of the $T$-linear resistivity is considerably more
stable. At low doping, where $\rho\sim T/x$ is expected, panel (b) shows
$xA$. The one-hole results are nearly independent of system size, even as
the doping changes from $x=1/16$ to $1/20$, as expected for
$\rho\sim T/x$. Thus, while the residual resistivity is progressively
suppressed with increasing size, the $T$-linear slope retains the expected
$1/x$ scaling. This stability suggests that the slope is an intrinsic
property of the semi-quantum regime rather than a finite-size artifact.
The two-hole result exhibits a larger correction, consistent with the
stronger size sensitivity seen in panel (a), but remains qualitatively
consistent with the dilute one-hole results.

A separate finite-size feature appears in the low-frequency optical
conductivity. We characterize the small downturn as $\omega\rightarrow0$
by the position of the low-frequency maximum of $\sigma(\omega)$, denoted
$\omega^\ast$. As shown in panel (c), $\omega^\ast/T\lesssim1$ for
$T\gtrsim0.08t$, encompassing most of the semi-quantum regime.
The downturn is therefore confined to a frequency scale well below the
thermal scale over the regime of interest. This separation of scales,
together with the agreement between the different extrapolation procedures
discussed above, shows that the extracted dc conductivity is insensitive
to this narrow low-frequency feature. Again, the two-hole sector shows
larger low-temperature corrections while following the same overall trend.

Finally, panel (d) shows the fraction of spectral weight not contained in the regular part of the optical conductivity,
\begin{equation}
f_D =
1-\frac{\displaystyle\int_{0^+}^\infty
d\omega\,\sigma(\omega)}
{(\pi/2)|K|},
\end{equation}
i.e., the Drude-weight fraction, which in the present context is a finite-size contribution that decreases systematically with increasing system size, remaining below $20\%$ over the regime of interest, consistent with its expected vanishing in the thermodynamic limit. Here $(\pi/2)|K|$ is the total optical weight on the half-axis $\omega\geq0$, with $K$ the thermodynamic kinetic energy per site. A finite zero-frequency contribution is a well-known artifact of optical-conductivity calculations on finite clusters with twisted boundary conditions \cite{Kohn1964,Poilblanc1991,ZotosNaefPrelovsek1997}: the many-body spectrum remains sensitive to flux threaded through the finite torus, giving a finite charge stiffness and hence a $\delta(\omega)$ peak, physically corresponding to persistent currents.
Note that this zero-frequency contribution does not enter our determination of $\sigma_{\rm dc}$, which is obtained by extrapolating the regular conductivity using data at $\omega\gtrsim\omega^\ast$, thereby excluding both the spurious low-frequency downturn and the $\omega=0$ Drude contribution. The missing weight in panel (d) therefore does not correspond to unresolved finite-frequency transitions contaminating the dc limit; moreover, its systematic decrease with system size indicates that this weight redistributes, in the thermodynamic limit, to frequencies captured by the regular part. Finite-size effects relevant to $\sigma_{\rm dc}$ are instead assessed in the main text through system-size dependence, boundary-condition comparisons, and the variance over twists.

\subsection{Optical lineshape diagnostics}
\label{sec:lineshape}

\begin{figure}[t!]
\centering
\includegraphics[width=\textwidth]{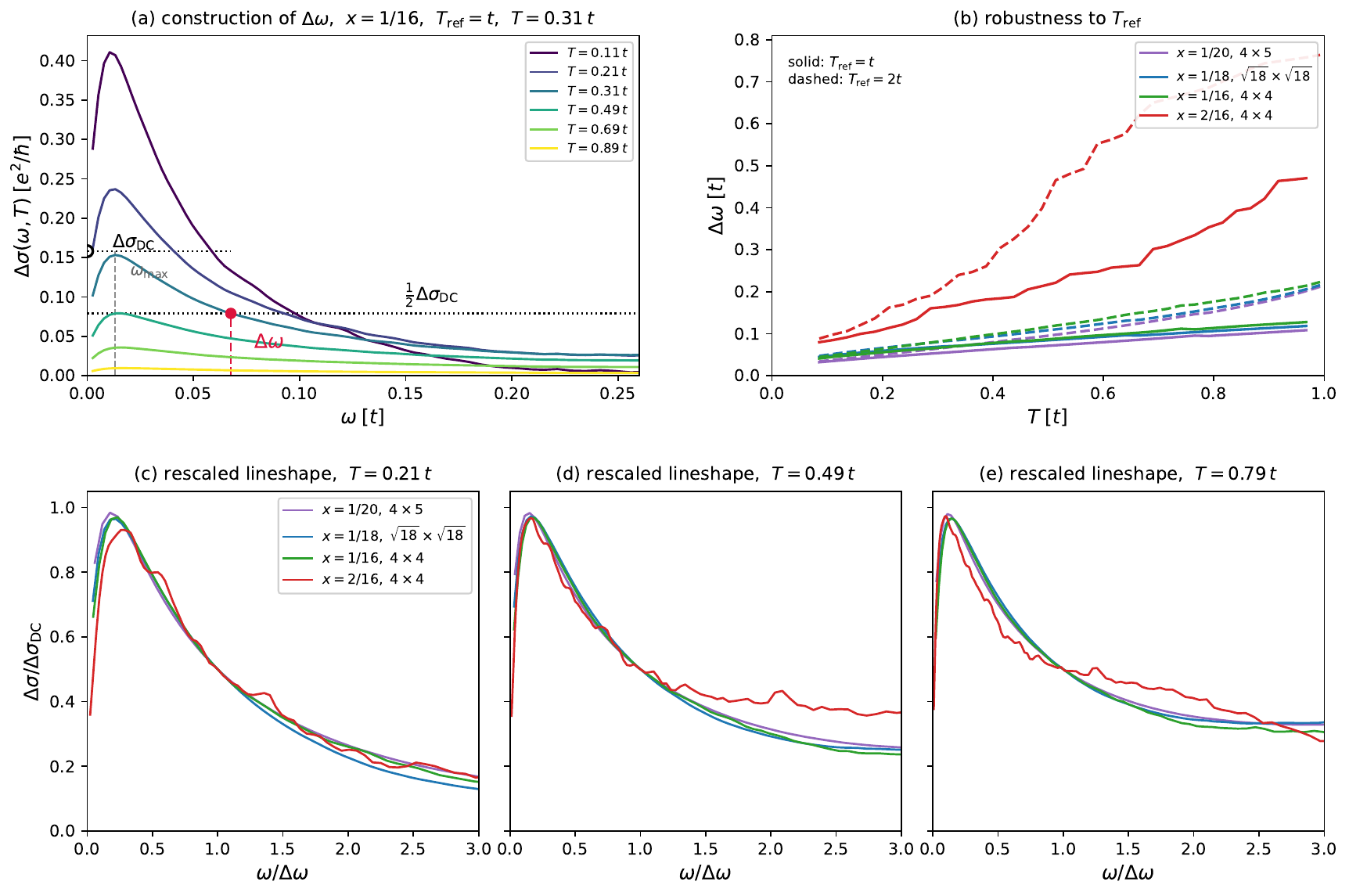}
\caption{\textbf{Characterization of the low-frequency optical feature.}
(A) Construction of the high-frequency half-maximum edge $\Delta\omega$ from
$\Delta\sigma(\omega,T)=\sigma(\omega,T)-\sigma(\omega,T_{\rm ref})$
for $x=1/16$ and $T_{\rm ref}=t$. The marked points indicate
$\omega_{\max}$ and the high-frequency half-maximum crossing
$\Delta\omega$.
(B) Dependence on the reference temperature. Solid and dashed curves show
the raw $\Delta\omega(T)$ obtained using $T_{\rm ref}=t$ and $2t$,
respectively, for the four systems shown in Fig.~\ref{fig:optical conductivity}. The $x=2/16$ values are
shown without the factor of $1/2$ used for visual comparison in
Fig.~\ref{fig:optical conductivity}F.
(C--E) Rescaled low-frequency lineshapes,
$\Delta\sigma(\omega,T)/\Delta\sigma_{\rm dc}(T)$ versus
$\omega/\Delta\omega(T)$, at the indicated representative temperatures.}
\label{fig:optical_lineshape_supp}
\end{figure}

To characterize the narrowing of the low-frequency feature without assuming a
specific lineshape, we define
$\Delta\sigma(\omega,T)=\sigma(\omega,T)-\sigma(\omega,T_{\rm ref})$.
We characterize its frequency extent by the high-frequency half-maximum edge
$\Delta\omega$, defined by
\[
\Delta\sigma(\Delta\omega,T)
=
\tfrac{1}{2}\Delta\sigma_{\rm dc}(T),
\qquad
\Delta\omega>\omega_{\max},
\]
where $\omega_{\max}$ is the position of the maximum of
$\Delta\sigma(\omega,T)$ [Fig.~\ref{fig:optical_lineshape_supp}A]. In the main text we use
$T_{\rm ref}=t$; the resulting $\Delta\omega(T)$ is shown in Fig.~\ref{fig:optical conductivity}F.

We test the dependence of this construction on the choice of reference
temperature in Fig.~\ref{fig:optical_lineshape_supp}B. Solid and dashed curves show the raw
$\Delta\omega(T)$ obtained using $T_{\rm ref}=t$ and $2t$, respectively.
Changing $T_{\rm ref}$ shifts the absolute values somewhat while leaving the
qualitative temperature dependence essentially unchanged. Unlike in Fig.~\ref{fig:optical conductivity}F, the $x=2/16$ values are
shown here without the factor of $1/2$, so that the absolute frequency scale
can be compared directly. The one-hole clusters exhibit an approximately
linear decrease of $\Delta\omega$ upon cooling, whereas the two-hole result
has a larger characteristic frequency scale and a qualitatively different
temperature dependence, as discussed in the main text.

As a further check that $\Delta\omega$ meaningfully characterizes the
low-frequency lineshape, Figs.~\ref{fig:optical_lineshape_supp}C--E show
$\Delta\sigma(\omega,T)/\Delta\sigma_{\rm dc}(T)$ versus
$\omega/\Delta\omega(T)$ at representative temperatures. The peak height and
high-frequency half-maximum point coincide by construction. Away from these
normalization points, the one-hole curves exhibit an approximate lineshape
collapse over a broader frequency range, while the two-hole lineshape shows
larger deviations, consistent with its qualitatively distinct behavior.

\subsection{Ioffe--Regel convention}
\label{sec:IR}

For reference we state the convention used for the Ioffe--Regel line in the
main text.  For a two-dimensional Drude metal,
\begin{equation}
\sigma = \frac{e^2 n \tau}{m} = \frac{e^2}{2\pi\hbar}\, k_F \ell .
\end{equation}
The validity of the Boltzmann transport equation (for Fermi liquids or non-Fermi liquids \cite{prange_transport_1964}) requires the mean free path to exceed the
Fermi wavelength, $\ell \gtrsim \lambda_F = 2\pi/k_F$, i.e.,
$k_F\ell/2\pi \gtrsim 1$, which translates into
\begin{equation}
\sigma_{\mathrm{IR}} = \frac{e^2}{\hbar}
\qquad\Longleftrightarrow\qquad
\rho_{\mathrm{IR}} = \frac{\hbar}{e^2}.
\label{eq:IR}
\end{equation}
Resistivities $\rho \gtrsim \hbar/e^2$ therefore signal transport beyond the
Boltzmann transport regime and in particular do not host electronic quasiparticles. Note that this criterion is distinct from
the Mott criterion $\ell > a$~\cite{mott1972}, which is violated only at
correspondingly larger resistivities.

\section{Resistivity in a bubble approximation}
\label{sec:bubble_cond}

Here we provide a limited examination of the importance of vertex
corrections to the resistivity by comparing the full Kubo result with the
``bubble'' conductivity constructed from the single-particle
spectral function.  We restrict this comparison to the $4\times4$ cluster
with one hole, $x=1/16$.  We compute the exact canonical finite-$T$
\textit{spin-summed} spectral function of the projected electrons
\begin{equation}
\tilde c_{\bm k\sigma}
=
\frac{1}{\sqrt N}\sum_j e^{-i\bm k\cdot\bm r_j}
c_{j\sigma}(1-n_{j\bar\sigma})
\end{equation}
by full diagonalization of the $N_e-1$, $N_e$, and $N_e+1$ sectors,
\begin{equation}
\begin{split}
A(\bm k,\omega,T)
= \sum_{\sigma,n,m}\frac{e^{-E_n^{(N_e)}/T}}{Z_{N_e}}
\Big[&
|\langle m^{(N_e-1)}|\tilde c_{\bm k\sigma}|n^{(N_e)}\rangle|^2
\delta_\gamma\!\left[\omega-(E_n^{(N_e)}-E_m^{(N_e-1)})\right]
\\[-2pt]
+&
|\langle m^{(N_e+1)}|\tilde c^\dagger_{\bm k\sigma}|n^{(N_e)}\rangle|^2
\delta_\gamma\!\left[\omega-(E_m^{(N_e+1)}-E_n^{(N_e)})\right]
\Big].
\end{split}
\label{eq:akw}
\end{equation}
Here $\delta_\gamma$ is a Lorentzian broadening.  Frequencies are not
shifted by the chemical potential, so that the Fermi level occurs at
$\omega=\mu(T)$.  The projected spectral function obeys
$\int d\omega\,A(\bm k,\omega,T)=2-n=1+x$, giving $17/16$ for the
cluster considered here.

From $A(\bm k,\omega,T)$ we evaluate the dc bubble conductivity
\cite{Deng2013},
\begin{equation}
\sigma^{\rm bub}_{x}(T)
=
\frac{\pi}{N}\sum_{\bm k}
\left(v^x_{\bm k}\right)^2
\int d\omega\,
\left(-\frac{\partial f(\omega-\mu)}{\partial\omega}\right)
\frac{A(\bm k,\omega,T)^2}{2},
\label{eq:bubble}
\end{equation}
where $v^x_{\bm k}=2t\sin\tilde{k}_x$.  The factor $1/2$ follows from
$\sum_\sigma A_\sigma^2=A^2/2$ for the spin-summed spectral function in
the spin-symmetric state.

Because this comparison is not the primary focus of the present work and
carries a substantially higher computational cost, we restrict the
calculation to four boundary twists and a coarser frequency grid.  This
provides meaningful results for $T\gtrsim0.2t$, sufficient for the limited
purpose here of assessing the role of vertex corrections.  Since
$A(\bm k,\omega,T)$ is evaluated canonically whereas
Eq.~(\ref{eq:bubble}) contains a grand-canonical Fermi window, for each
twist we determine $\mu$ from
\begin{equation}
\sum_{\bm k}\int d\omega\,
f(\omega-\mu)A(\bm k,\omega,T)=N_e
\label{eq:mu_number}
\end{equation}
with $f$ the Fermi-Dirac distribution function.
We restrict the comparison to temperatures for which the thermally sampled
window of width $\sim T$ around $\mu$ is sufficiently broad that
finite-size ambiguities in the precise location of the chemical potential
are unimportant.

\begin{figure}[t!]
\centering
\includegraphics[width=\textwidth]{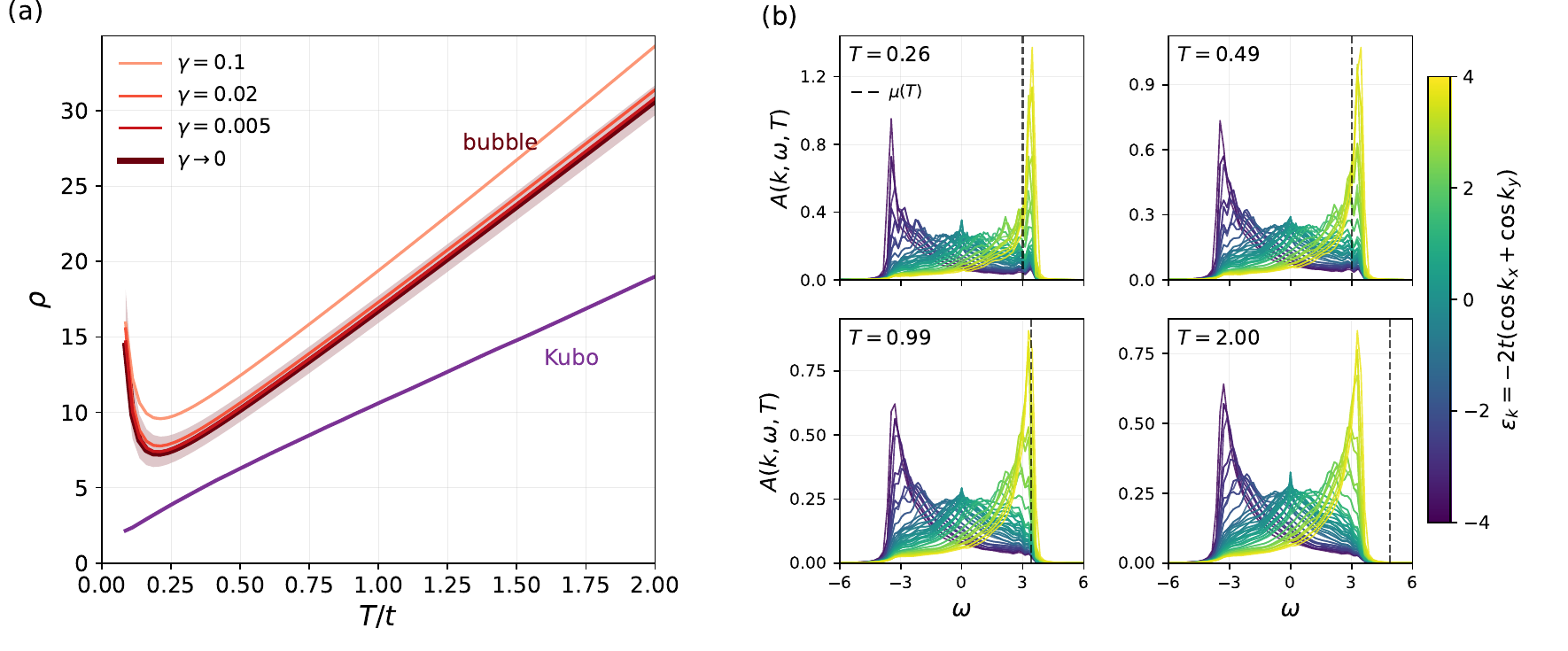}
\caption{\textbf{Single-particle spectral function and bubble
resistivity.}
(\textbf{a}) Full Kubo resistivity and the vertex-free bubble resistivity
$\rho^{\rm bub}$ obtained from Eq.~(\ref{eq:bubble}) for the
$4\times4$ one-hole cluster.
(\textbf{b}) Spin-summed spectral function
$A(\bm k,\omega,T)$ entering Eq.~(\ref{eq:bubble}), shown for
$\gamma=0.02t$ at four representative temperatures.  Curves correspond
to the $64$ twist-shifted momenta obtained by pooling the $16$ momenta
from each of the four boundary twists used in the bubble calculation;
the curves are not averaged over twists. Lines are colored by the
bare-band energy
$\varepsilon_{\bm k}=-2t(\cos k_x+\cos k_y)$ of the corresponding
twist-shifted momentum.  Frequencies are not shifted by the chemical
potential, and dashed vertical lines indicate $\mu(T)$ determined from
Eq.~(\ref{eq:mu_number}).}
\label{fig:bubble}
\end{figure}

Figure~\ref{fig:bubble} shows the resulting bubble
resistivity together with the full Kubo result, alongside representative
single-particle spectral functions entering Eq.~(\ref{eq:bubble}).  The
bubble resistivity,
$\rho^{\rm bub}=1/\langle\sigma^{\rm bub}\rangle_{\bm\theta}$, differs
quantitatively from the full Kubo resistivity of the finite cluster, while
exhibiting a similar temperature dependence.  Over $0.3\lesssim T/t\leq2$,
\begin{equation}
\frac{\rho^{\rm bub}(T)}{\rho^{\rm Kubo}(T)}
\approx 1.6 ,
\end{equation}
with little detectable temperature dependence.  Thus we find that vertex corrections
are important for quantitatively capturing the finite-cluster
resistivity, although the bubble shows similar $T$-scaling over this range.  The importance of such corrections
is consistent with the quantitative discrepancy between
16-site FTLM and single-site DMFT resistivities reported for the doped
Hubbard model in Ref.~\cite{Brown2019}.  We defer a systematic study of
these effects and a
more detailed investigation of the single-particle spectral function, which is directly accessible in cold-atom Fermi--Hubbard systems through
ARPES-type measurements \cite{brown2020arpes}, to
future work.

\section{Details on the hole-spin-bond correlator}
\label{sec:triplecorr}

\subsection{Definition and normalization}

Recall the connected hole-spin-bond correlator defined in Eq.~\eqref{eq:nSS} is
\begin{equation}
C_h(\mathbf R,\mathbf e;T)
=
\left\langle
\rho_h(\mathbf 0)\,
\mathbf S_{\mathbf R-\mathbf e/2}\!\cdot\!
\mathbf S_{\mathbf R+\mathbf e/2}
\right\rangle
-
\left\langle \rho_h(\mathbf 0)\right\rangle
\left\langle
\mathbf S_{\mathbf R-\mathbf e/2}\!\cdot\!
\mathbf S_{\mathbf R+\mathbf e/2}
\right\rangle ,
\qquad
\left\langle \rho_h(\mathbf 0)\right\rangle
=
x=\frac{1}{N},
\label{eq:Chdef}
\end{equation}
where $\mathbf R$ is the shortest displacement, accounting for periodic
boundary conditions, from the origin to the bond center, and
$\mathbf e=\hat{\mathbf x},\hat{\mathbf y}$ specifies the bond orientation.
We denote by $C_h(R,T)$ the value of $C_h$ at $R=|\mathbf R|$.
We normalize throughout by
\begin{equation}
C_0=\frac{x^2}{2},
\label{eq:C0}
\end{equation}
the $T\to0$ value of \eqref{eq:Chdef} on any bond away from the hole,
derived below. The four $R=1/2$ bonds incident on the conditioned hole site
are called \emph{incident}.

\subsection{A sum rule}

For brevity we denote $b=(i,i+\mathbf e)$ the nearest-neighbor
bond and
$B=\sum_b \mathbf S_i\!\cdot\!\mathbf S_{i+\mathbf e}$ the total
nearest-neighbor bond operator, with the sum running over all $2N$ bonds.
With exactly one hole,
$\sum_{\mathbf r}\rho_h(\mathbf r)=1$ as an operator identity, while both
$B$ and the thermal state are translation invariant. Hence
\begin{equation}
\langle \rho_h(\mathbf 0) B\rangle
=
\frac{1}{N}\sum_{\mathbf r}
\langle \rho_h(\mathbf r) B\rangle
=
\frac{1}{N}\langle B\rangle
=
x\langle B\rangle .
\end{equation}
Summing \eqref{eq:Chdef} over all bonds therefore gives
\begin{equation}
\sum_b C_h(b;T)
=
\langle \rho_h(\mathbf 0)B\rangle
-
x\langle B\rangle
=
0,
\label{eq:sumrule}
\end{equation}
where $C_h(b;T)$ is shorthand for
$C_h(\mathbf R_b,\mathbf e_b;T)$.

\subsection{The Nagaoka ground state}

As $T\to0$, the $U=\infty$ one-hole ground state is the Nagaoka
ferromagnet~\cite{Nagaoka1966}, with maximal total spin
$S=(N-1)/2$ and a uniformly delocalized hole, so that
$\langle\rho_h(\mathbf r)\rangle=x$ for every $\mathbf r$.
Two elementary observations then fix \eqref{eq:Chdef} completely.

First, conditioned on the hole sitting at the origin, all remaining spins
belong to the maximally polarized $S=(N-1)/2$ multiplet. Hence, on any bond
that avoids the origin,
\begin{equation}
\left\langle
\mathbf S_i\!\cdot\!\mathbf S_{i+\mathbf e}
\right\rangle_h
=
\frac14,
\end{equation}
i.e.
\begin{equation}
\left\langle
\rho_h(\mathbf 0)\,
\mathbf S_i\!\cdot\!\mathbf S_{i+\mathbf e}
\right\rangle
=
\frac{x}{4}.
\end{equation}
On an incident bond one endpoint is the conditioned hole site and therefore
carries no spin, so the joint expectation vanishes identically as an operator identity.

Second, without conditioning on the hole position, either endpoint of a given
bond is empty with probability $x$, so the probability that both endpoints
are occupied is $1-2x$. In the fully polarized state this gives
\begin{equation}
\left\langle
\mathbf S_i\!\cdot\!\mathbf S_{i+\mathbf e}
\right\rangle
=
\frac14(1-2x).
\end{equation}

Combining the two, the $T\to0$ values are
\begin{equation}
C_h^{\,\mathrm{non\text{-}inc}}
=
\frac{x}{4}
-
x\,\frac{1-2x}{4}
=
\frac{x^2}{2}
=
C_0,
\qquad
\frac{C_h(R=1/2,T\to0)}{C_0}
=
\frac{-x(1-2x)/4}{x^2/2}
=
-\frac{1-2x}{2x},
\label{eq:nagaoka}
\end{equation}
which is $-8$ for $N=18$. The first of these is Eq.~\eqref{eq:C0}: in the
fully polarized state the normalized correlator on every non-incident bond is
exactly $+1$, independently of $N$ and of its distance from the hole.
The two limits in \eqref{eq:nagaoka} are not independent: with 4 incident and
$2N-4$ non-incident bonds, the sum rule \eqref{eq:sumrule} requires
\begin{equation}
4\times\left[-\frac{N-2}{2}\right]
+
(2N-4)\times1
=
0,
\end{equation}
which holds identically.

The incident shell is special because it contains no conditional local bond
information. Since the conditional part vanishes there identically,
\eqref{eq:Chdef} reduces to minus the disconnected piece alone. Defining the
orientation-averaged nearest-neighbor spin correlation by
\begin{equation}
\left\langle
\mathbf S_i\!\cdot\!\mathbf S_{i+\mathbf e}
\right\rangle_{\rm nn}
\equiv
\frac{1}{2}
\sum_{\mathbf e=\hat{\mathbf x},\hat{\mathbf y}}
\left\langle
\mathbf S_i\!\cdot\!\mathbf S_{i+\mathbf e}
\right\rangle ,
\end{equation}
the shell-averaged incident correlator obeys the exact identity
\begin{equation}
\frac{C_h(R=1/2,T)}{C_0}
=
-\frac{2}{x}\,
\left\langle
\mathbf S_i\!\cdot\!\mathbf S_{i+\mathbf e}
\right\rangle_{\rm nn}
\qquad\text{at all }T .
\label{eq:incident}
\end{equation}
The deep blue column at $R=1/2$ in Fig.~\ref{fig:cage}A is therefore not a
statement about the local bonds surrounding a conditioned hole; it directly
tracks the global nearest-neighbor spin correlation, up to rescaling.

\subsection{Formation of the ferromagnetic texture}

\begin{figure}[t!]
\centering
\includegraphics[width=\textwidth]{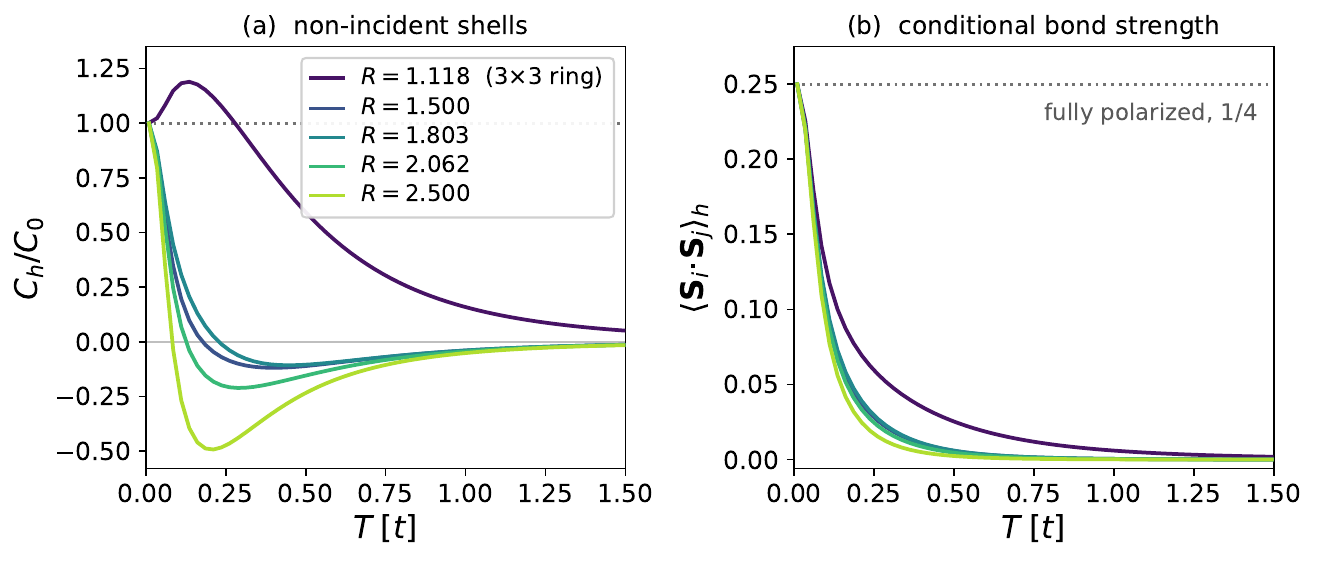}
\caption{\textbf{Shell-resolved hole-spin-bond correlator, 18-site cluster,
$x=1/18$, untwisted.}  Bonds are grouped by the distance $R$ from the hole to the bond
center; the incident shell $R=1/2$, fixed exactly by Eq.~\eqref{eq:incident}, is not
shown.  (\textbf{a})~$C_h/C_0$, Eq.~\eqref{eq:Chdef} normalized by its Nagaoka value
$C_0=x^2/2$ (dotted).  (\textbf{b})~The conditional bond strength
$\langle \mathbf S_i\!\cdot\!\mathbf S_j\rangle_h =
\langle \rho_h(0)\,\mathbf S_i\!\cdot\!\mathbf S_j\rangle/x$, same shells and colors,
approaching the fully polarized $1/4$ (dotted).}
\label{fig:triplecorr}
\end{figure}

Fig.~\ref{fig:triplecorr}A shows the five non-incident shells. All approach
the common Nagaoka value $+1$ as $T\to0$, but their evolution is quite
different. The $R=\sqrt{5}/2$ shell, consisting of the eight bonds forming
the $3\times3$ ``spin cage'' around the hole, is the only one that overshoots this
value, peaking at intermediate $T$ inside the semi-quantum regime.
The more distant shells instead become negative at intermediate temperature.
By the sum rule \eqref{eq:sumrule}, any enhancement of the connected
correlation near the hole must be compensated elsewhere.

Since $C_h$ also contains a temperature-dependent disconnected contribution,
the overshoot above $+1$ does not mean that the local bond correlation exceeds
that of the fully polarized state. To characterize the local texture more
directly, we consider the conditional bond correlation
\begin{equation}
\left\langle
\mathbf S_i\!\cdot\!\mathbf S_{i+\mathbf e}
\right\rangle_h
\equiv
\frac{
\left\langle
\rho_h(\mathbf 0)\,
\mathbf S_i\!\cdot\!\mathbf S_{i+\mathbf e}
\right\rangle
}{x},
\end{equation}
shown in Fig.~\ref{fig:triplecorr}B. It vanishes identically on the incident
shell, approaches the fully polarized value $1/4$ on all other shells as
$T\to0$, and is largest on the $R=\sqrt{5}/2$ ring. Its magnitude, however,
remains unsaturated down to very low temperatures. In other words, below
$\Tq$, the spins near the hole are clearly correlated ferromagnetically, but
the local texture remains fluctuating and rather than fully polarized.

\section{High-temperature expansions}
\label{sec:HTE}

The dashed high-temperature curves in the main-text figures
(Figs.~\ref{fig:rho_kappa_D}B and \ref{fig:thermo}) follow from the
leading high-temperature expansion (HTE) of the free energy, which we
summarize here.  At $T \gg t$ the canonical partition function of a cluster
of $V$ sites with $N_e$ electrons is expanded as
\begin{equation}
Z_{N_e} = \Omega_{N_e}
\left( 1 - \beta \langle H\rangle_0
        + \tfrac{\beta^2}{2} \langle H^2 \rangle_0 + \dots \right),
\qquad
\Omega_{N_e} = \binom{V}{N_e} 2^{N_e},
\label{eq:ZHTE}
\end{equation}
where $\langle\cdot\rangle_0$ denotes the infinite-temperature average and $\Omega_{N_e}$ counts the constrained configurations.
Since $H$ is purely off-diagonal in the occupation basis,
$\langle H\rangle_0 = 0$; moreover $\langle H^3\rangle_0 = 0$ on the bipartite clusters (the $4\times4$
torus and the 18-site Betts cluster), because their shortest closed loops are
the four-bond plaquettes. The one caveat concerns the staggered ($s=1$) torus and the $4\times5$ torus are not bipartite. Their shortest
odd loops wind the boundary and have length $L_y+s=5$ and $L_y=5$,
respectively.  Odd moments therefore first enter at order $\beta^5$, well
beyond the order kept below, and on the bipartite clusters they vanish
identically.  The second moment
counts the allowed directed nearest-neighbor hops:
a hop $i\to j$ requires site $i$ occupied and site $j$ empty, which at fixed
$N_e$ occurs with probability $N_e(V-N_e)/[V(V-1)]$; with $4V$ directed
nearest-neighbor links this gives
\begin{equation}
a_2 \equiv \langle H^2\rangle_0
= 4 t^2\, \frac{N_e\,(V-N_e)}{V-1}
\;\xrightarrow{\;V\to\infty\;}\; 4t^2 n(1-n)\, V ,
\label{eq:a2}
\end{equation}
with $n = N_e/V = 1-x$.  From
$F = -T\ln Z = -T \ln\Omega_{N_e} - a_2/(2T) + \dots$ one obtains the
leading high-temperature forms, per site,
\begin{align}
s(T) &= s_\infty - \frac{2t^2}{T^2}\, n(1-n),
\qquad
s_\infty = -n\ln (n/2) - (1-n)\ln(1-n),
\label{eq:sHTE}\\[2pt]
c_V(T) &= \frac{4t^2}{T^2}\, n(1-n),
\label{eq:cvHTE}\\[2pt]
\chic^{-1}(T) &= \frac{T}{n(1-n)} + \frac{4t^2}{T}.
\label{eq:kinvHTE}
\end{align}
For comparison with the finite clusters at fixed particle number we use the
finite-$V$ expressions, obtained from Eq.~(\ref{eq:ZHTE}) without taking the
thermodynamic limit:
\begin{align}
s &= \frac{1}{V}\ln \Omega_{N_e}
   - \frac{2t^2}{T^2}\, \frac{N_e (V-N_e)}{V(V-1)},
\label{eq:sHTEfin}\\[2pt]
c_V &= \frac{4t^2}{T^2}\, \frac{N_e (V-N_e)}{V(V-1)},
\label{eq:cvHTEfin}\\[2pt]
\chic^{-1} &= V T\,
\ln\!\left[ \frac{(N_e+1)(V-N_e+1)}{N_e\,(V-N_e)} \right]
+ \frac{4Vt^2}{(V-1)\,T},
\label{eq:kinvHTEfin}
\end{align}
where the first term of Eq.~(\ref{eq:kinvHTEfin}) is obtained by inserting
$F_{N_e} = -T\ln\Omega_{N_e}$ into the definition Eq.~(\ref{eq:chic}).  These finite-cluster forms are the dashed curves shown
in the main text.

For the spin sector, independent spin-$1/2$ moments give the Curie law with
the main-text normalization, $T\chis = N_e/(4N) = (1-x)/4$.  In the opposite
(Nagaoka) limit the ground-state multiplet has total spin
$S_{\mathrm{tot}} = N_e/2$; averaging $S_z^2$ over the degenerate multiplet
gives $\langle S_z^2\rangle = S_{\mathrm{tot}}(S_{\mathrm{tot}}+1)/3$, hence
\begin{equation}
T\chis \;=\; \frac{N_e (N_e+2)}{12\,N} .
\label{eq:nagaoka_chis}
\end{equation}
The departure of each observable from its high-temperature form defines the
crossover scale $\Tq$ operationally; as expected for a crossover (rather
than a transition), the resulting scale varies somewhat between observables,
$\Tq \approx 0.5$--$0.8\,t$.

\section{Nagaoka ferromagnetism and spin stiffness}
\label{sec:stiffness}

The lower boundary of the semi-quantum regime is set by the coherence scale
$\Tcoh$ associated with the ferromagnetic correlations of the low-doping
ground state~\cite{Nagaoka1966,LiuYaoBergWhiteKivelson2012}.  Here we estimate the associated
spin stiffness for a single hole and recall how it sets $\Tcoh \sim xt$.

For one hole on a cluster of $N$ sites, the ground state is the fully
polarized Nagaoka ferromagnet, and the hole occupies the top of the
single-particle band at $\mathbf k_0 = (\pi,\pi)$: with all $\mathbf k$
states but $\mathbf k_0$ filled by spin-aligned electrons,
\begin{equation}
E_0 = \sum_{\mathbf k \neq \mathbf k_0} \epsilon(\mathbf k)
    = -\epsilon(\mathbf k_0) = -4t,
\qquad
\epsilon(\mathbf k) = -2t\,(\cos k_x + \cos k_y).
\end{equation}
To extract the stiffness of the ferromagnet we impose a uniform spin twist
of total angle $\phi$ across the $x$~direction, i.e., a local spin rotation
by $n_x \alpha$ with $\alpha = \phi/L$.  The local ``up'' direction then
rotates from site to site, and the overlap of neighboring spinors reduces
the effective hopping of the polarized band,
\begin{equation}
t \;\to\; \big\langle \uparrow_{\mathbf r} \big| \uparrow_{\mathbf r + \hat x} \big\rangle\, t
= \cos\!\Big( \frac{\phi}{2L} \Big)\, t ,
\end{equation}
for $x$~bonds only.  Keeping the hole at $\mathbf k_0$, the ground-state
energy shift is
\begin{equation}
\Delta E(\phi) = 2t \left[ 1 - \cos\!\Big( \frac{\phi}{2L} \Big) \right]
\approx \frac{t\,\phi^2}{4 L^2}.
\end{equation}
Matching to the continuum stiffness definition in $d=2$,
$\Delta E / L^d = (\rho_s/2)\, (\phi/L)^2$, gives
\begin{equation}
\rho_s = \frac{x\,t}{2}, \qquad x = \frac{1}{N},
\label{eq:rhos}
\end{equation}
i.e., a stiffness of order $xt$, carried by the mobile holes.
At any finite temperature the two-dimensional ferromagnet is disordered, but
with an exponentially growing correlation length,
\begin{equation}
\xi_{\mathrm{FM}}(T) \sim \exp\!\left( \frac{2\pi \rho_s}{T} \right),
\label{eq:xiFM}
\end{equation}
so that for $T \lesssim \Tcoh \sim xt$ the holes effectively move through an
ordered spin background as coherent spinless fermions.

\section{Variational bound for the caging free energy}
\label{sec:caging}

The caging construction of the main text is controlled by the thermodynamic
variational principle.  Let $H_{\mathrm{tr}}$ be obtained from $H$ by cutting
all hopping bonds that cross the boundary of a chosen cage region, with the
hole inside the cage and the exterior half filled.  The Gibbs--Bogoliubov
inequality, $F \le F_{\mathrm{tr}} + \langle H - H_{\mathrm{tr}}\rangle_{\mathrm{tr}}$,
immediately gives $F \le F_{\mathrm{tr}}$: $H_{\mathrm{tr}}$ conserves the
particle numbers of the cage and of the exterior separately, every term of
$H - H_{\mathrm{tr}}$ transfers a particle across the boundary, and hence
$\langle H - H_{\mathrm{tr}}\rangle_{\mathrm{tr}} = 0$.  Moreover, at
$U=\infty$ the half-filled exterior has no charge dynamics, so the trial free
energy is additive,
\begin{equation}
F_{\mathrm{tr}}(\Ncage, T)
= F_{\mathrm{cage}}(\Ncage, T) - (N - \Ncage)\, T \ln 2 ,
\label{eq:Ftr}
\end{equation}
with $F_{\mathrm{cage}}$ the free energy of the isolated cage (an
open-boundary cluster of $\Ncage$ sites containing one hole), obtained by
exact diagonalization of the cage alone.
Throughout this work $\DFcage \equiv F_{\mathrm{tr}} - F \ge 0$ denotes the
caging free-energy cost.  At high temperatures it approaches the
positional-entropy cost of confining the hole,
$\DFcage \to T\ln(N/\Ncage)$, which follows from the exact $T\to\infty$
state counts $N\,2^{N-1}$ and $\Ncage\,2^{N-1}$ of the full and trial
systems, respectively.  We therefore consider the
residual cost
\begin{equation}
\Delta\widetilde F_{\rm cage}(\Ncage,T) \;\equiv\;
\DFcage(\Ncage,T) - T\ln(N/\Ncage),
\label{eq:DFcage}
\end{equation}
which measures the confinement cost beyond positional entropy.  The figures
show the dimensionless ratio $\Delta\widetilde F_{\rm cage}/T$: it vanishes
at high temperature, where the free energy is local; is of order unity
through much of the semi-quantum window, so that confinement on the cage
scale costs of order the thermal scale; and grows rapidly on cooling toward
$\Tcoh$, as discussed in the main text.

\subsection{Cage geometries and effective cage length}
\begin{figure}[t!]
\centering
\includegraphics{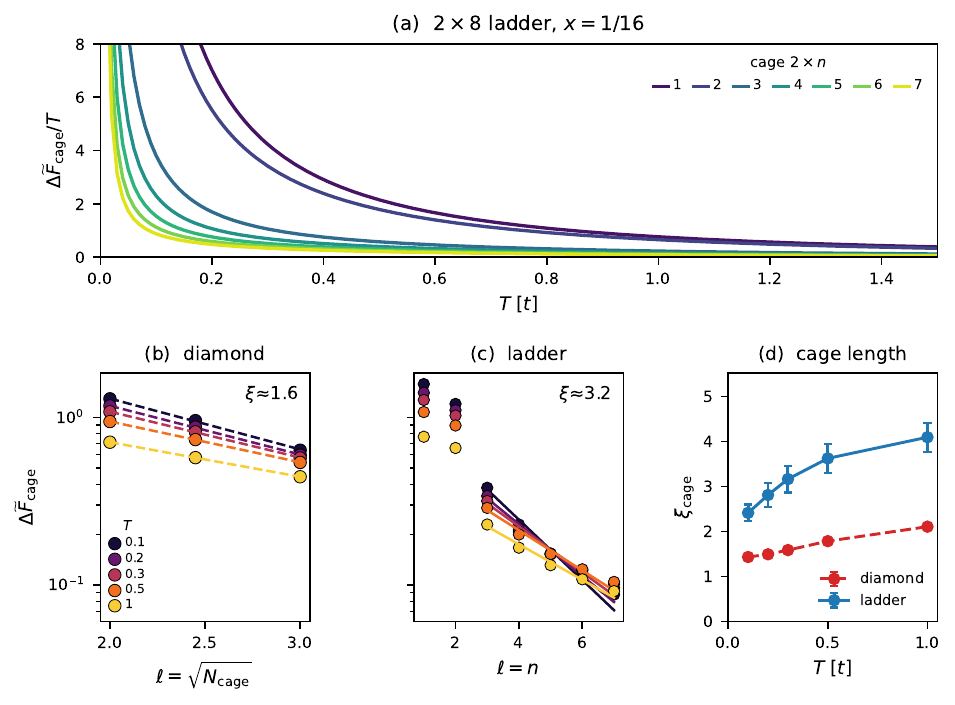}
\caption{\textbf{Cage confinement cost, cage-size scaling, and cage length.}
(\textbf{a})~Residual confinement cost $\Delta \widetilde{F}_{\rm cage}/T$
[Eq.~(\ref{eq:DFcage})] versus temperature for the $2\times n$ cages
($n = 1,\dots,7$) on the $2\times8$ ladder ($x=1/16$).  Bonds crossing the
cage boundary are cut; the hole is confined to the cage and the exterior
remains half filled.  (The corresponding cost for the 18-site diamond is
shown in Fig.~\ref{fig:cage}C.)  Throughout the semi-quantum window the cost remains
order the thermal scale, $\ \Delta \widetilde{F}_{\rm cage}/T \sim 1$, growing rapidly only on
approaching $\Tcoh$.  (\textbf{b},\textbf{c})~Cage-size dependence of the
confinement cost: $\Delta \widetilde{F}_{\rm cage}$ versus cage linear size $\ell$ on a logarithmic
scale, with the exponential fits of Eq.~(\ref{eq:xicagefit})
[$\ell = \sqrt{N_{\rm cage}}$ for the 18-site diamond cages $2\times2$,
$3\times2$, $3\times3$; $\ell = n$ for the $2\times n$ ladder cages], at
temperatures $T = 0.1$--$1.5\,t$.  (\textbf{d})~Fitted cage length
$\xicage(T)$ for both geometries.}
\label{cageSM}
\end{figure}

Figure~\ref{cageSM}(a) shows the residual confinement cost of the
$2\times8$ ladder with a single hole ($x=1/16$), for cages $2\times n$ with
$n = 1,\dots,7$.  This quasi-one-dimensional geometry complements the
two-dimensional 18-site probe of the main text and, admitting a longer series
of cage sizes, resolves the cage-size dependence.  Its confinement cost
behaves as on the 18-site cluster, of order of the thermal scales across the semi-quantum window and
rising on cooling toward $\Tcoh$, showing that effective caging, with a
comparable cage length, is not an artifact of the specific two-dimensional
cluster.

Over the accessible range of cage sizes, the dependence on cage size is  approximately exponential,
\begin{equation}
\Delta \widetilde{F}_{\rm cage}(T, \ell) \;=\; A(T)\, e^{-\ell/\xicage(T)},
\qquad
\ell =
\begin{cases}
\sqrt{\Ncage} & \text{(2D cages)},\\[2pt]
n & \text{($2\times n$ ladder cages)},
\end{cases}
\label{eq:xicagefit}
\end{equation}
which defines an effective cage length $\xicage(T)$
[Fig.~\ref{cageSM}(b,c,d)].  The fitted length is short and weakly
temperature dependent across the semi-quantum window,
$\xicage \approx 1.5$ lattice spacings on the 18-site cluster and
$\approx 2.8$ on the ladder.  A short, nearly $T$-independent $\xicage$
supports the interpretation of charge motion in the semi-quantum regime as
quasi-local and controlled by a finite spin environment around the hole rather
than by delocalization across the system.  On cooling toward $\Tcoh$ the
dimensionless confinement cost $\Delta \widetilde{F}_{\rm cage}/T$ rises sharply
[Fig.~\ref{cageSM}(a)], signaling the breakdown of the caging
description as ferromagnetic coherence sets in and the hole delocalizes.

\end{document}